\documentclass[reprint,aps, superscriptaddress,nofootinbib]{revtex4-1}

\usepackage{graphicx}
\usepackage{tikz}
\usepackage{dcolumn}% Align table columns on decimal point
\usepackage{bm}% bold math
\usepackage{color}
\usepackage{amsmath}
\usepackage{amssymb}
\usepackage{amsfonts}
\usepackage{amsthm}
\usepackage{subfig}
\usepackage{MnSymbol}
 \definecolor{ForestGreen}{rgb}{0.133, 0.545, 0.133}
 
\begin{document}

\title{Machine learning topological defect formation: When are the defects  made?}

\date{\today}

\author{Fumika Suzuki}
\email{fsuzuki@icepp.s.u-tokyo.ac.jp}
\affiliation{International Center for Elementary Particle Physics (ICEPP), The University of Tokyo, 7-3-1 Hongo, Bunkyo-ku, Tokyo
113-0033, Japan}

 \affiliation{%
Theoretical Division, Los Alamos National Laboratory, Los Alamos, New Mexico 87545, USA
}

 \affiliation{%
Center for Nonlinear Studies, Los Alamos National Laboratory, Los Alamos, New Mexico 87545, USA
}

\author{Ying Wai Li}

 \affiliation{%
Computer, Computational, and Statistical Sciences Division, Los Alamos National Laboratory,
Los Alamos, NM 87545, USA
}

\author{Wojciech H. Zurek}

 \affiliation{%
Theoretical Division, Los Alamos National Laboratory, Los Alamos, New Mexico 87545, USA
}

  \begin{abstract}
Topological defects that form in a nonequilibrium second-order phase transition are presumably seeded by fluctuations of the order parameter in the vicinity of the critical point. Motivated by this conjecture that underlies the Kibble-Zurek mechanism (KZM), we investigate whether machine learning (ML) can anticipate their locations from the fluctuations in the ``impulse regime'', the time interval when the evolution of the order parameter cannot keep up with the conditions imposed by the quench. Going beyond the conventional KZM focus on defect density, we show that a recurrent neural network can  predict locations of topological defects from short-time dynamical data deep within the impulse regime. We thus demonstrate that defects are sown in the immediate vicinity of the critical point. The seeds of defects—fluctuations imprinted on the evolving order parameter---are exponentially small near the critical point, but become amplified by the evolution and play a dominant role in breaking symmetry. This is before the freezeout time $\hat t$ that concludes the impulse regime, and well before the order parameter assumes its final symmetry-broken configuration. Furthermore, we find that the predictive power of the ML also exhibits power-law scaling consistent with KZM.
\end{abstract}

\maketitle

%\tableofcontents

\begin{figure*}
\centering
{%
\includegraphics[clip,width=1.5\columnwidth]{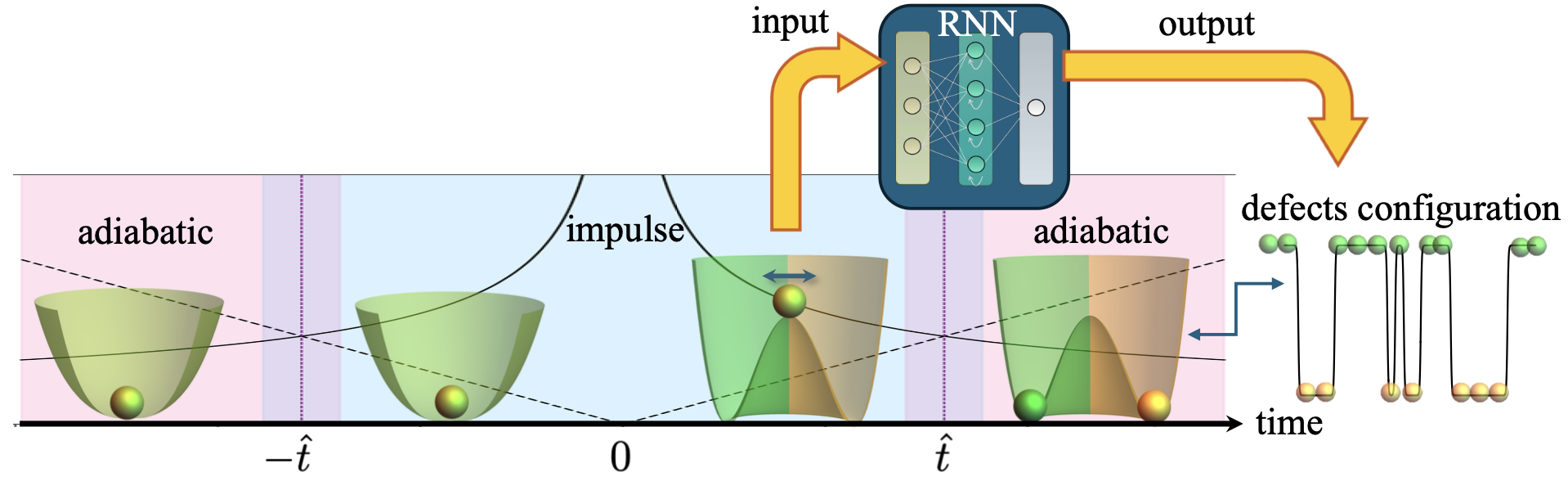} 
}
\caption{\textbf{Machine learning  prediction of topological defect configurations based on KZM.}  The sequential inputs consist of short time series representing the time evolution of the order parameter during a second-order phase transition. We use an RNN to predict the final configuration of the order parameter.}
\label{fig1}
\end{figure*}

\begin{figure}
\centering

{%
\includegraphics[clip,width=0.6\columnwidth]{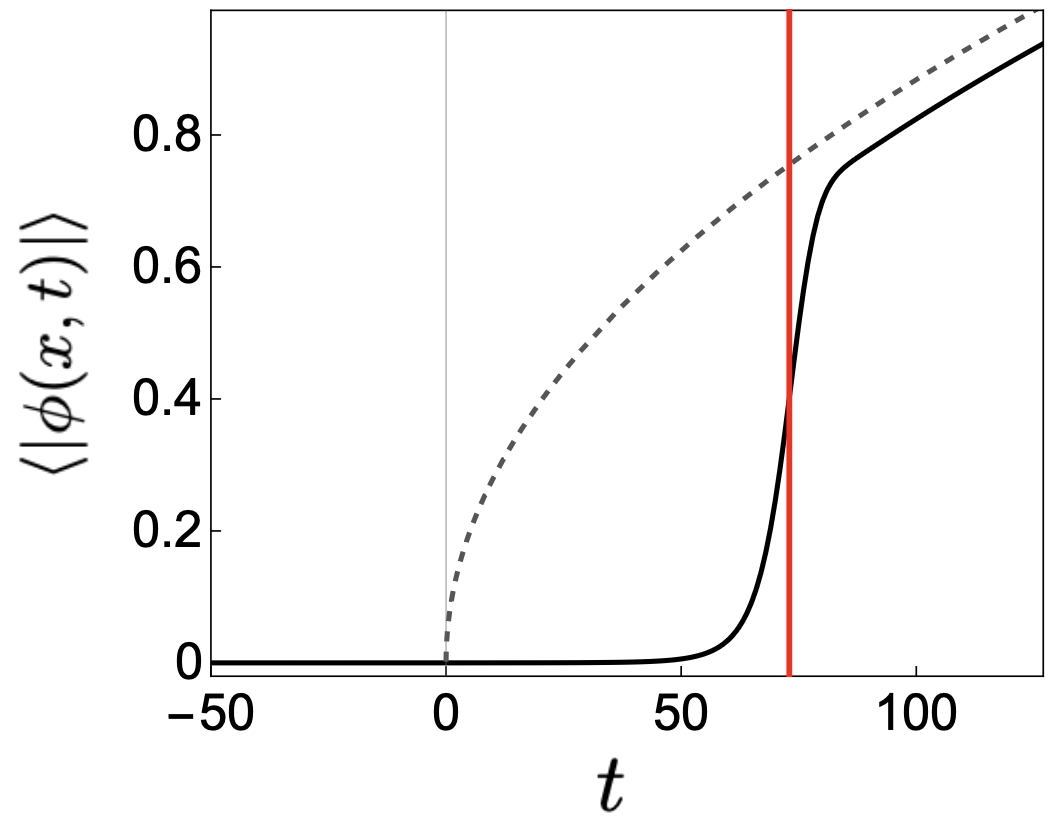} 
}
\caption{\textbf{Dynamics of the order parameter.} The time evolution of $\langle |\phi (x,t)|\rangle$ where spatial position $x$ is  averaged over (solid black line). As predicted by the AIA picture, $\phi$ remains frozen at the location of the old minimum ($\phi=0$) for a while even after the spontaneous symmetry breaking at $t = 0$. Then it catches up to the location of the emerging minima at $\phi = \pm \sqrt{\epsilon}$ around the freeze-out time $\hat{t}$.  The red line indicates the time at which $\langle |\frac{d}{dt }\phi (x,t)|\rangle$ reaches its maximum, our estimate of $\hat t$. The dashed grey line represents the location of the emerging minima $|\phi|=\sqrt{\epsilon}$ following spontaneous symmetry breaking at $t = 0$. $\tau_Q=128$.
} 
\label{fig2}
\end{figure}

The Kibble-Zurek mechanism (KZM) combines the inevitability of topological defect formation in cosmological phase transitions noted by Kibble \cite{kzm1,kzm2} with the theory \cite{kzm3,kzm4,kzm5} that relates their density to the critical slowing down and, hence, to the universality class of the second order phase transition. 
KZM has found wide-ranging applications  across diverse fields, from cosmology \cite{kzm1,kzm2,kzm3,kzm4,kzm5} to condensed matter physics \cite{kzm6, issac, lin, spinice, condensed,condensed2,condensed3,condensed4,condensed5,condensed6} and even quantum computing \cite{damski2005simplest, polkovnikov2005universal, zurek2005dynamics, dziarmaga2005dynamics, dziarmaga2010dynamics, rams2019symmetry, schmitt2022quantum, zhang2025kibble, bayocboc2025persistent, dziarmaga2022coherent, gardas2018defects, qc3,qc,qc2,qc4,xu2018dynamic}.

The key insight \cite{kzm3,kzm4,kzm5} that leads to KZM scaling is the realization that, near the critical point of the second order phase transition, critical slowing down will result in an ``impulse’’ time interval $[-\hat{t}, +\hat{t}] $ where the order parameter is too sluggish to adjust to the potential that is changing faster than its relaxation time $\tau$. Thus, while outside this interval the order parameter can be in approximate equilibrium, within the interval $[-\hat{t}, +\hat{t}]$ its evolution ``cannot keep up''. The system is then incapable of adjusting to the external changes in the evolving potential imposed by the quench, which results in creation of defects. Fluctuations imparted after $ -\hat{t}$ seed topological defects that germinate at $ +\hat{t}$ in ways that depend on the nature of the system.

The symmetry-breaking dynamics underlying KZM were recently analyzed within a deconstruction framework, which separates the temporal and spatial evolution of the order parameter, providing a more complete understanding of the dynamics of symmetry breaking and explaining the physical origin of the conventional adiabatic–impulse–adiabatic (AIA) picture \cite{suzuki2025deconstructing}. The results motivate the conjecture that machine learning (ML) should be able to predict the defect {\it locations} from the early-time evolution of the order parameter within the impulse regime, before it settles into the symmetry-broken minima (Fig. \ref{fig1}). ML has been widely used to classify and analyze phases and phase transitions in classical and quantum systems \cite{ML,phaseML,phaseML2,phaseML3,phaseML4,phaseML5,phaseML6,phaseML7,phaseML8,phaseML9,quantumML,quantumML2,quantumML3, scheurer2020unsupervised, kaming2021unsupervised, che2026quantum}, as well as nonequilibrium dynamics \cite{noneq,noneq2,noneq3, donatella2023dynamics}. However, its application to the study of topological defect formation and dynamics remains relatively limited \cite{chowdhury2023topological, dierking2025machine}. In this paper, we employ ML as a tool to pinpoint when the defect seeds appear. We shall see that they are sown already in the immediate vicinity of the critical point, when the expectation value of the evolving order parameter is very small, orders of magnitude smaller than outside of the impulse regime. 
We demonstrate this using a recurrent neural network (RNN). 
 %This is before the freezeout time $+\hat t$, and well before the order parameter assumes its final symmetry-broken configuration. 
This is surprising, as at that time the amplitude of the order parameter is exponentially suppressed \cite{suzuki2025deconstructing}, which makes it impossible to identify defect seeds with the ``naked eye''. Remarkably, the ML results suggest that limited information about the defect locations is present even {\it before} the critical point is traversed. Thus, ML can go beyond the conventional KZM: while KZM predicts universal scaling of defect densities, RNN can learn to predict, from the early-time evolution that results in symmetry breaking, locations of topological defects. We further show that ML predictability  exhibit universal behavior rooted in the power-law scaling of KZM. \\

RNNs are a type of artificial neural network specifically designed to process sequential or time series data  \cite{RNN, RNN2, RNN3}. They incorporate recurrent connections, allowing the output from one time step to be fed back as input at the next. This feedback mechanism enables RNNs to capture temporal dependencies within the data, facilitating the development of ML models capable of predicting future states based on sequential inputs. Here, the sequential inputs are short intervals of the time evolution of the order parameter during a second-order phase transition. RNN is employed to predict the locations of topological defects once the order parameter settles into the emerging minima of the 
%spontaneously 
symmetry-breaking potential. We  show that the ML model can accurately predict the locations of topological defects—i.e., ``kinks'', the points where the order parameter changes sign in our one-dimensional Landau-Ginzburg model—using short-time sequences of evolving order parameter in the impulse regime as input.\\

The training data is prepared as follows \cite{GitHubCode}. The formation of topological defects (kinks) during a second-order phase transition can be described by the Langevin equation in (1+1) dimensions for a real scalar field $\phi$ representing the order parameter that evolves under a Landau-Ginzburg potential  \cite{laguna,laguna2,suzuki,suzuki2025deconstructing, jacek}:
\begin{eqnarray}\label{eq}
\ddot{\phi}+\eta \dot{\phi} -\partial_{xx}\phi +\partial_{\phi} V(\phi)=\vartheta (x,t)
\end{eqnarray}
where $V(\phi)= (\phi^4-2\epsilon \phi^2 )/8$. As dimensionless distance from the critical point $\epsilon (t)=t/\tau_Q$ (where $\tau_Q$ is the quench timescale) changes from negative to positive, spontaneous symmetry breaking occurs, leading to the formation of topological defects. 
%$\vartheta$ represents 
The noise is characterized by its correlation function,  $\langle \vartheta (x,t), \vartheta  (x',t')\rangle = 2\eta \theta \delta (x'-x)\delta (t'-t)$. We set the damping constant to $\eta=1$  (i.e., evolution is overdamped), and noise temperature to $\theta=10^{-8}$. Note that different temperatures, e.g., $\theta=10^{-2}$, yield similar results (see Supplemental Material). The spatial computational domain has size \( L = 512 \), discretized into \( 1024 \) grid points with $dx = 0.5$ and periodic boundary conditions.

\begin{figure}
\centering
{%
\includegraphics[clip,width=0.85\columnwidth]{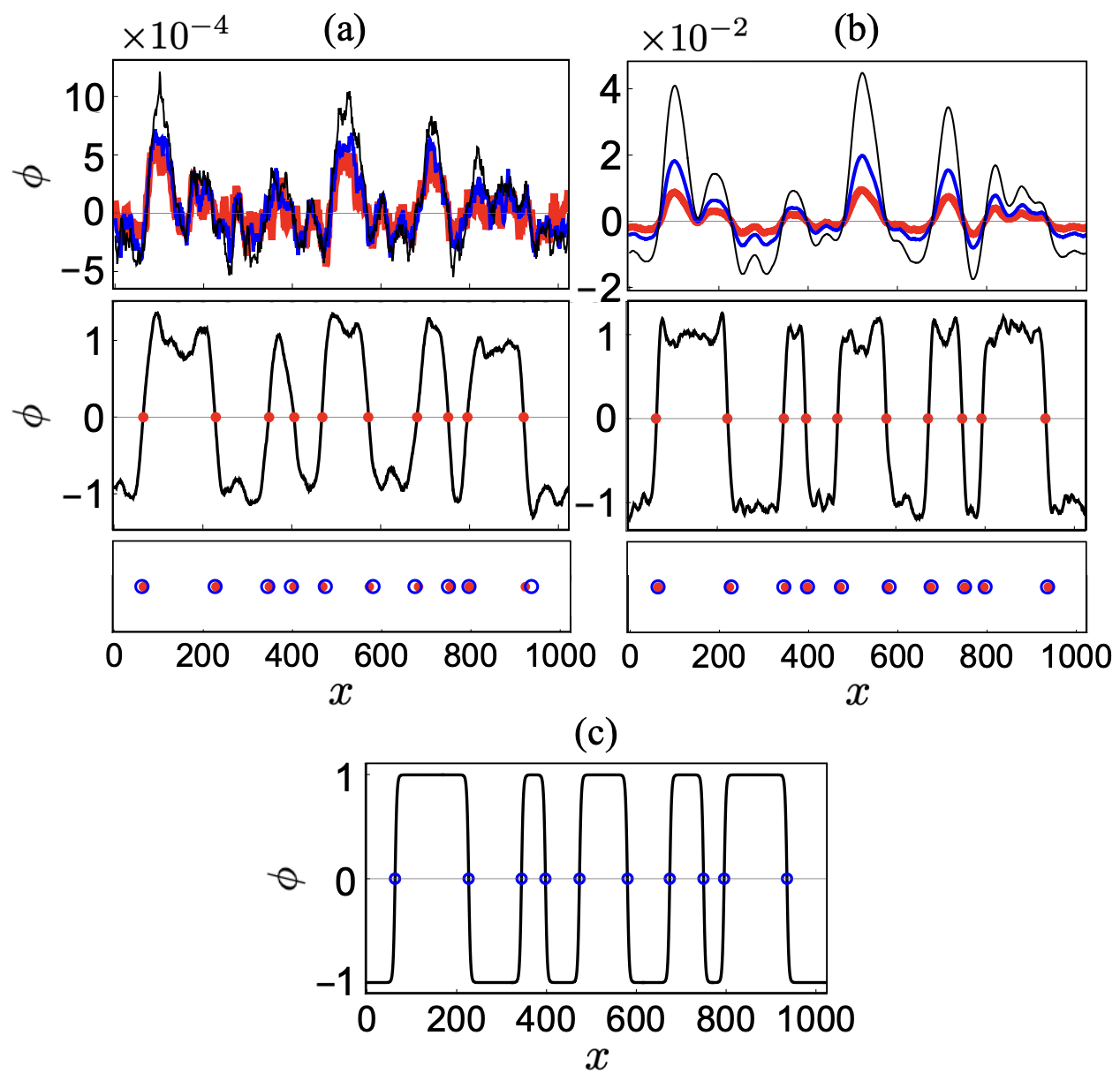} 
}
\caption{\textbf{Examples of the prediction process performed by RNN.}  (a, b): The upper panels show the input data, the middle panels show the predicted final topological defect configurations, and the lower panels compare the predicted (red filled circles) and true (blue open circles) locations of the topological defects.  The input data consist of time series of $\phi$, where each series spans from $t-5$ to $t+5$ with a time step $\Delta t=1$. $t = 30$ $(\epsilon=0.23)$ for (a), and $t = 50$ $(\epsilon =0.39)$ for (b). The upper panels show $\phi$ at $t-5$ ($\epsilon -0.039$; thick red line), $t$ ($\epsilon$; blue line), and $t+5$  ($\epsilon +0.039$; thin black line) as representative examples. (c): True final configuration of $\phi$. $\tau_Q=128$.} 
\label{fig3}
\end{figure}

\begin{figure*}
\centering
{%
\includegraphics[clip,width=1.4\columnwidth]{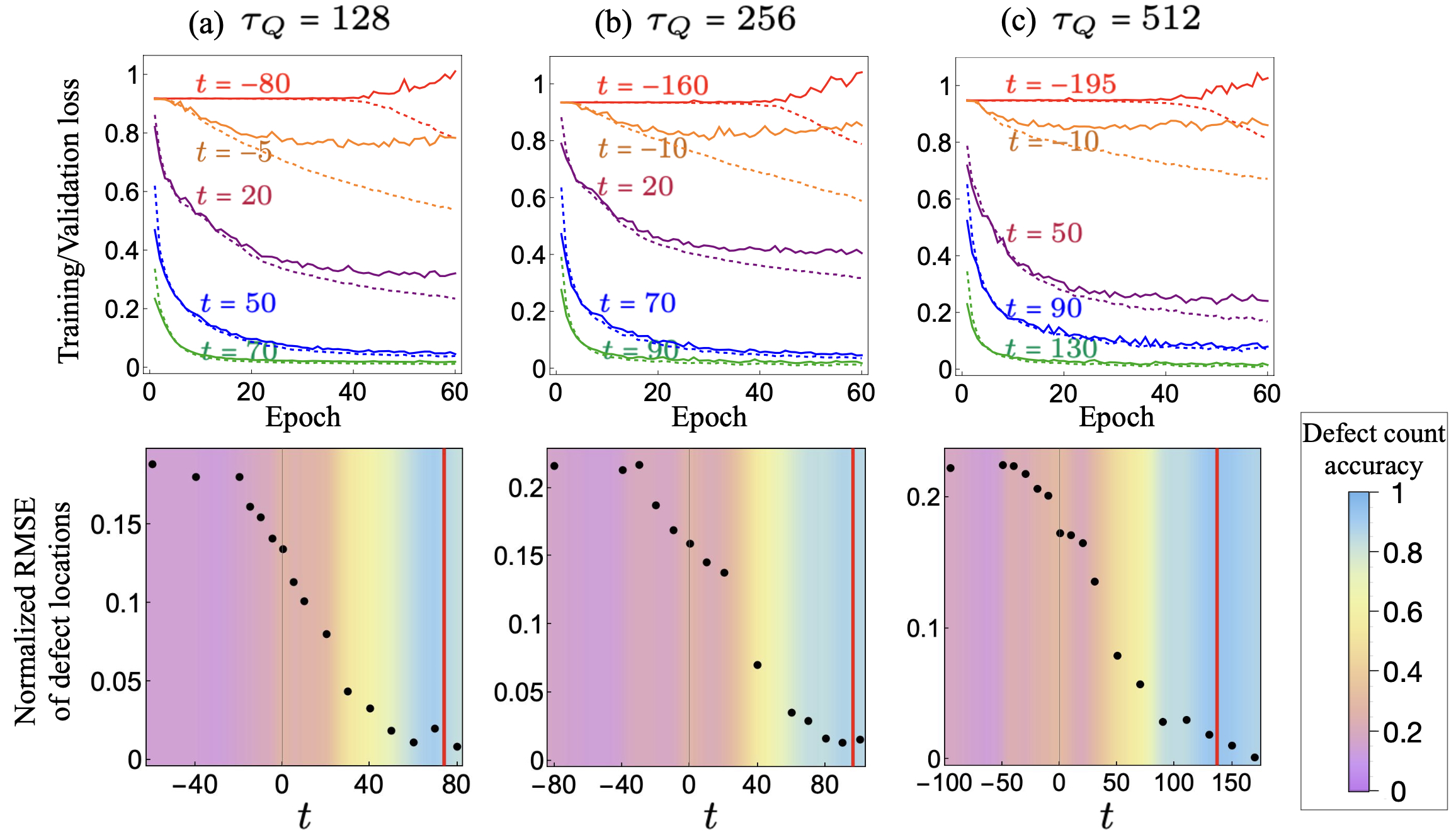} 
}
\caption{\textbf{Machine learning  for predicting defect locations.}  $\tau_Q = 128$ for (a), $256$ for (b), and $512$ for (c). Upper panels:  Training (dashed) and validation loss (solid) plotted as a function of epoch, based on input time series sampled at different times. (a) $t=-80, -5, 20, 50,70$, (b) $t=-160,-10, 20,70,90$ and (c) $t=-195,-10, 50,90,130$  for the red, orange, purple, blue, and green lines, respectively (from top to bottom).  Lower panels: Normalized RMSE between the predicted and true defect locations, evaluated only when the model correctly predicted the number of defects. The defect count accuracy (i.e., the fraction of cases where the number of defects is correctly predicted) is represented by the color plot.  Those are evaluated using 100 testing samples after training the ML model for 60 epochs on 3000 samples. The red vertical lines indicate the time at which $\langle |\frac{d}{dt }\phi (x,t)|\rangle$ reaches its maximum after the phase transition at $t = 0$.}
\label{fig4}
\end{figure*}

We numerically computed the time evolution of the order parameter using Eq. (\ref{eq}) in which $\epsilon$ changes from –1.5 to 1, and obtained training data by repeating the process for multiple random noise realizations (Supplemental Material).

 Fig. \ref{fig2} shows  the time evolution of $\langle |\phi (x,t)|\rangle$ where  $x$ is  averaged over, and the quench timescale $\tau_Q=128$. As predicted by the AIA picture, $\phi$ fluctuates  around the old symmetric minimum at $\phi = 0$ for a while even after the critical point is crossed at $t=0$ (i.e., the order parameter  ceases to follow its equilibrium value since it is unable to adjust to the externally imposed change in the potential due to the divergence of the relaxation time).  Then the order parameter catches up with the emerging new minima at $\phi = \pm\sqrt{\epsilon}$ represented by the dashed grey line around the freeze-out time $+\hat{t}$. The red line indicates the time at which $\langle |\frac{d}{dt }\phi (x,t)|\rangle$ reaches its maximum after the phase transition at $t = 0$. The freeze-out of the defect configuration occurs around this point \cite{ suzuki2025deconstructing}. 

Fig.~3 illustrates examples of the prediction process performed by the RNN for quench timescale \(\tau_Q = 128\).
 The input data consists of time series $\phi (x,t')$, where each series spans from $t'=t-5$ to $t+5$ with a time step $\Delta t= 1$ (11 sequential data points).  An RNN is  employed to predict the final configuration of the order parameter \(\phi\) at \(t = \tau_Q\) with \(\epsilon = 1\). The predicted locations of topological defects (i.e., points where \(\phi = 0\)) are then compared with the true final defect locations (Supplemental Material)
It shows that, even from the short-time evolution of small fluctuations in \(\phi\) during the impulse regime—between the critical point at \(t = 0\) and before the freeze-out time \(+\hat{t}\)—the model can predict the final locations of topological defects with notable accuracy. This result can be understood in light of the recent deconstruction of KZM \cite{suzuki2025deconstructing}. In this framework, the temporal and spatial evolution of the order parameter are analyzed separately, revealing how fluctuations imprinted near the critical point are selectively amplified and become encoded in the long-wavelength structure of the order parameter. These long-wavelength structures subsequently evolve into the final domains and determine the defect locations, whereas short-wavelength fluctuations are largely transient and relax away. The RNN may therefore learn to identify the long-wavelength structures that survive and evolve into the final domain pattern while disregarding transient fluctuations that are later erased by the dynamics.

\begin{figure}
\centering
{%
\includegraphics[clip,width=0.75\columnwidth]{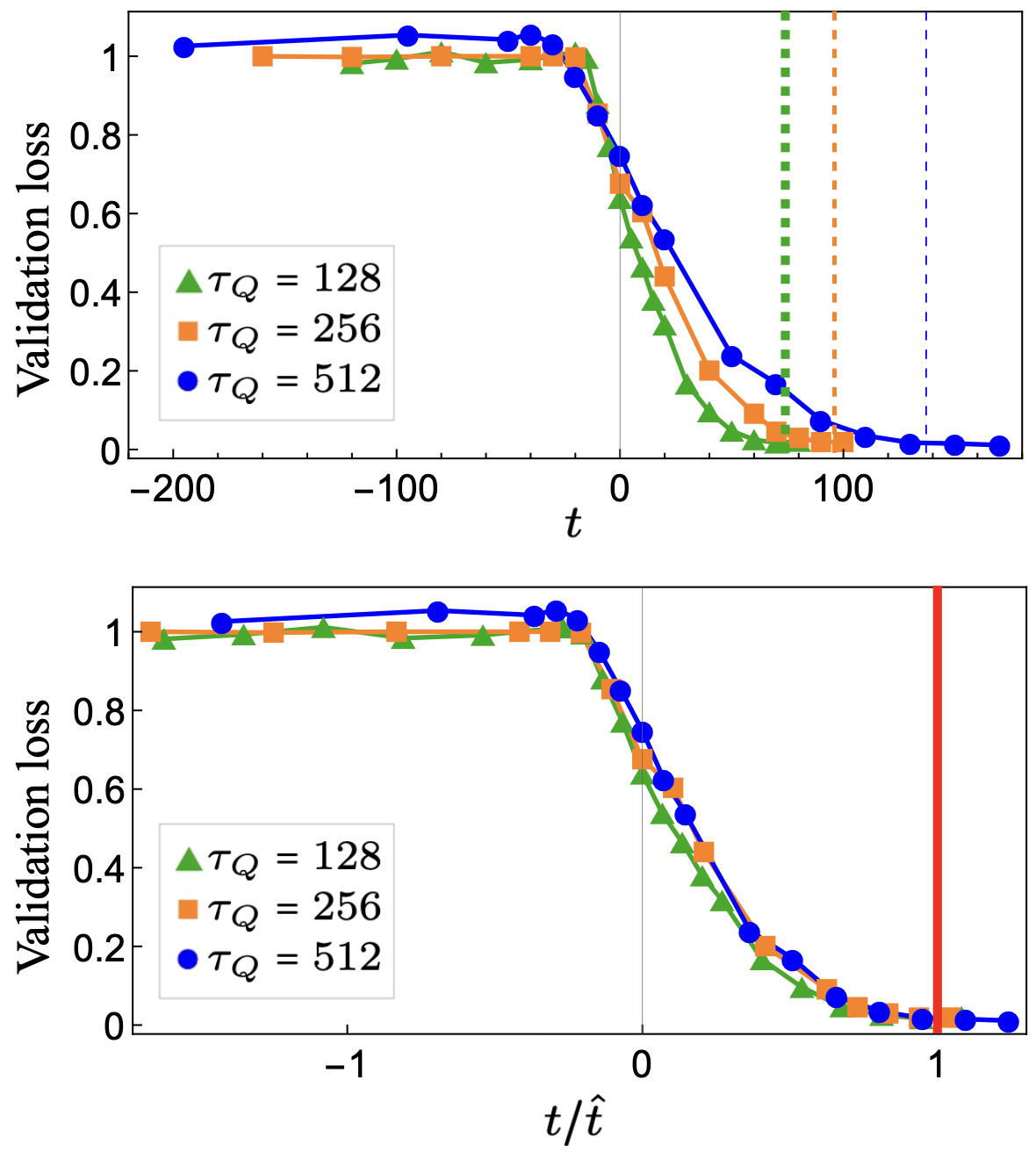} 
}
\caption{\textbf{ML predictability follows the power law scaling implied by KZM.} Validation loss for different quench timescale $\tau_Q$ achieved after 60 training epochs.  The upper panel is plotted as a function of time \( t \), while the lower panel is plotted as a function of the rescaled time $t/\hat{t}$, the KZM freeze-out time: Validation loss curves coincide after the rescaling. Vertical lines in both panels represent the corresponding freeze-out time for each \( \tau_Q \).
}
\label{fig5}
\end{figure}

The upper panels in Fig. \ref{fig4}  show the training loss (dashed) and validation loss (solid) plotted as a function of epoch.  The loss function is the mean squared error (MSE) between the predicted and true $\phi$ at the time when $t=\tau_Q$ with $\epsilon = 1$. The training and validation losses correspond to this quantity evaluated on the training and validation datasets, respectively. It can be seen that the ML model can predict part of the final topological defect configuration even from input data obtained shortly before the critical point is traversed at $t=0$. After the critical point, ML begins to  predict the final topological defect configuration more effectively, even though the inputs are from a time series well before the freeze-out time $+\hat{t}$.  The lower panels show  the normalized root mean squared error (RMSE) between the predicted and true defect locations, where RMSE is divided by the  size of the spatial
computational domain $L$. Both of them are evaluated using 100 testing samples after training the ML model for 60 epochs on 3000  samples of the time evolution of the order parameter. The RMSE of the defect locations is computed only when the model correctly predicts the number of defects. The defect count accuracy (i.e., the fraction of cases where the number of defects is correctly predicted) is represented by the color plot. The figure demonstrates that even when the input data capture only a short period of the time evolution of small fluctuations in $\phi$ around $\phi=0$ during the impulse regime, the model can predict both the number and final locations of topological defects with high precision. This is consistent with KZM, where the adiabatic--impulse conjecture suggests that topological defects are seeded during the impulse regime, before the freeze-out time $+\hat{t}$, with $\hat{t} \propto \sqrt{\eta \tau_Q}$, when the order parameter $\phi$ begins to settle into the emerging new minima that mark spontaneous symmetry breaking \cite{kzm3,kzm4,kzm5}.

 The importance of the fluctuations imprinted on $\phi(x)$ in the immediate vicinity of the critical point (hence, the relative unimportance of those imparted at $t<0$, or even these from just before $\hat t$) may be surprising, but it is a consequence of  the attenuation--instantiation--amplification (aia) dynamics induced by  $V(\phi)= (\phi^4-2\epsilon \phi^2 )/8$: Before $t=0$ the quadratic term in $V(\phi)$ represents a restoring force, with a time-dependent harmonic oscillator-like ``spring constant’’ $-\epsilon=-t/\tau_Q$ that suppresses fluctuations of $\phi(x)$ (attenuation). This is why the pre-transition configuration from $t<0$ is erased. Near $t=0$, the restoring force vanishes as the potential becomes nearly flat, allowing noise to imprint fluctuations on $\phi(x)$ (instantiation). After $t=0$, as $-\epsilon<0$, the force changes sign (harmonic oscillator is now ``upside down’’) and acts as an amplifier (amplification). Perturbations of $\phi(x)$ imprinted by noise on $\phi(x)$ near $t=0$ become amplified more than these introduced later. This is why the configuration of the order parameter $\phi(x)$ near $t \approx 0$ matters more for defect locations than the noise just before $\hat t$. 
 % the plateau in Fig. 5 ends near $t=0$, although 
As our results indicate, noise that influenced $\phi(x)$ just before $t=0$ (where the harmonic potential is nearly flat) already begins to matter: This is when the harmonic potential ``flips'', initiating the amplification phase of symmetry breaking (see \cite{suzuki2025deconstructing}). The analytical solution derived in \cite{suzuki2025deconstructing} shows that fluctuations imprinted at this stage undergo the strongest amplification, whereas those introduced later have progressively less time to grow. This behavior is fully consistent with the conventional AIA picture, while providing a physical explanation of the dynamics that underlie it.

 Note that RMSE increases slightly around the red line. This is because, by then, some regions of $\phi$ begin to roll down toward the emerging minima. As a result, the input data can exhibit distinct types of fluctuations in $\phi$ that range from small values near $\phi = 0$ to larger values, where $\phi$ starts to roll down. This makes it difficult to analyze the data effectively, especially for the rapid quench timescales $\tau_Q$. \\

Finally, we demonstrate that ML predictability also follows the universal behavior captured by KZM. Fig. \ref{fig5} shows the validation loss achieved after 60 training epochs  for various quench timescales $\tau_Q$ (see the upper panels of Fig. \ref{fig4}).  The onset of predictability is best understood within the ``KZM deconstruction" paradigm \cite{suzuki2025deconstructing}: As discussed above, the pre-transition field configuration is essentially erased as crossing of the critical point at $t=0$ approaches. Therefore, it is of little consequence for the eventual defect locations. Consequently, the RNN cannot reliably predict the final defect pattern from the $t < 0$ configuration of $\phi(x)$. 
Once the system enters the near-critical regime, the noise resets the spatial structure of $\phi(x)$, and the subsequent evolution amplifies it. In particular, long-wavelength fluctuations that determine the final domains become effectively frozen, encoding information about the eventual defect configuration. 
We therefore interpret the rapid decrease in prediction error as a signature of the relevant information becoming accessible to the RNN. In this picture, the edge of the plateau is determined by the transition between perturbations that are erased and those that are amplified. Furthermore, it is known that the relevant timescale for the amplification process is the freeze-out time that follows power law scaling, $\hat{t} \propto \sqrt{\eta\tau_Q}$ in the overdamped regime for the Landau-Ginzburg model, Eq. (\ref{eq}) \cite{laguna2,suzuki,suzuki2025deconstructing}. In the rescaled time \( t/\hat{t} \) the plots for different \( \tau_Q \) exhibit nearly identical decaying behavior, as shown in the lower panels of Fig. \ref{fig5}. This observation aligns with the KZM scaling,  \cite{kzm3,kzm4,kzm5}, and especially the insights offered by its deconstruction \cite{suzuki2025deconstructing}.
The rescaling of time by \( \hat{t} \) leads to universal dynamics of the order parameter across different quench timescales, consequently resulting in the well-known universal power law for the density of topological defects as a function of the quench timescale
 \cite{kzm3, kzm4, kzm5, suzuki2025deconstructing}. We note that the same prediction behavior is observed with LSTM and Transformer architectures, indicating that it originates from the underlying physics of symmetry-breaking dynamics rather than from the specific RNN architecture (Supplemental Material).
 
 % \cite{= \exp{\frac {t^2} {{\hat t}^2}}.\\

The ability of the ML model to predict the final topological defect configuration presented in this paper is consistent with the Kibble–Zurek mechanism (KZM), but goes beyond its standard predictions by identifying when information about the final defect configuration becomes encoded in the order parameter. In this sense, our results provide support for the recent deconstruction of symmetry-breaking dynamics \cite{suzuki2025deconstructing}, which separates the temporal and spatial aspects of symmetry breaking and predicts that the information determining the final defect configuration is encoded already near the critical point.

\begin{acknowledgements}
We thank Kipton Barros, Avanish Mishra and Saryu Fensin for fruitful discussions. F.S. acknowledges support from the Los Alamos National Laboratory LDRD program under project number 20230049DR, the Center for Nonlinear Studies under project number 20250614CR-NLS and Office of Science. Y.W.L. acknowledges support from the Los Alamos National Laboratory LDRD program under project number 20240083DR. This work was supported in part by funding from Google.org.
\end{acknowledgements}

%\begin{appendices}

%\section{Section title of first appendix}\label{secA1}

%%=============================================%%
%% For submissions to Nature Portfolio Journals %%
%% please use the heading ``Extended Data''.   %%
%%=============================================%%

%%=============================================================%%
%% Sample for another appendix section			       %%
%%=============================================================%%

%% \section{Example of another appendix section}\label{secA2}%
%% Appendices may be used for helpful, supporting or essential material that would otherwise 
%% clutter, break up or be distracting to the text. Appendices can consist of sections, figures, 
%% tables and equations etc.

%\end{appendices}

%%===========================================================================================%%
%% If you are submitting to one of the Nature Portfolio journals, using the eJP submission   %%
%% system, please include the references within the manuscript file itself. You may do this  %%
%% by copying the reference list from your .bbl file, paste it into the main manuscript .tex %%
%% file, and delete the associated \verb+\bibliography+ commands.                            %%
%%===========================================================================================%%

\bibliography{sn-bibliography}% common bib file

%merlin.mbs apsrev4-1.bst 2010-07-25 4.21a (PWD, AO, DPC) hacked
%Control: key (0)
%Control: author (8) initials jnrlst
%Control: editor formatted (1) identically to author
%Control: production of article title (-1) disabled
%Control: page (0) single
%Control: year (1) truncated
%Control: production of eprint (0) enabled
\begin{thebibliography}{62}%
\makeatletter
\providecommand \@ifxundefined [1]{%
 \@ifx{#1\undefined}
}%
\providecommand \@ifnum [1]{%
 \ifnum #1\expandafter \@firstoftwo
 \else \expandafter \@secondoftwo
 \fi
}%
\providecommand \@ifx [1]{%
 \ifx #1\expandafter \@firstoftwo
 \else \expandafter \@secondoftwo
 \fi
}%
\providecommand \natexlab [1]{#1}%
\providecommand \enquote  [1]{``#1''}%
\providecommand \bibnamefont  [1]{#1}%
\providecommand \bibfnamefont [1]{#1}%
\providecommand \citenamefont [1]{#1}%
\providecommand \href@noop [0]{\@secondoftwo}%
\providecommand \href [0]{\begingroup \@sanitize@url \@href}%
\providecommand \@href[1]{\@@startlink{#1}\@@href}%
\providecommand \@@href[1]{\endgroup#1\@@endlink}%
\providecommand \@sanitize@url [0]{\catcode `\\12\catcode `\$12\catcode
  `\&12\catcode `\#12\catcode `\^12\catcode `\_12\catcode `\%12\relax}%
\providecommand \@@startlink[1]{}%
\providecommand \@@endlink[0]{}%
\providecommand \url  [0]{\begingroup\@sanitize@url \@url }%
\providecommand \@url [1]{\endgroup\@href {#1}{\urlprefix }}%
\providecommand \urlprefix  [0]{URL }%
\providecommand \Eprint [0]{\href }%
\providecommand \doibase [0]{http://dx.doi.org/}%
\providecommand \selectlanguage [0]{\@gobble}%
\providecommand \bibinfo  [0]{\@secondoftwo}%
\providecommand \bibfield  [0]{\@secondoftwo}%
\providecommand \translation [1]{[#1]}%
\providecommand \BibitemOpen [0]{}%
\providecommand \bibitemStop [0]{}%
\providecommand \bibitemNoStop [0]{.\EOS\space}%
\providecommand \EOS [0]{\spacefactor3000\relax}%
\providecommand \BibitemShut  [1]{\csname bibitem#1\endcsname}%
\let\auto@bib@innerbib\@empty
%</preamble>
\bibitem [{\citenamefont {Kibble}(1976)}]{kzm1}%
  \BibitemOpen
  \bibfield  {author} {\bibinfo {author} {\bibfnamefont {T.~W.}\ \bibnamefont
  {Kibble}},\ }\href@noop {} {\bibfield  {journal} {\bibinfo  {journal}
  {Journal of Physics A: Mathematical and General}\ }\textbf {\bibinfo {volume}
  {9}},\ \bibinfo {pages} {1387} (\bibinfo {year} {1976})}\BibitemShut
  {NoStop}%
\bibitem [{\citenamefont {Kibble}(1980)}]{kzm2}%
  \BibitemOpen
  \bibfield  {author} {\bibinfo {author} {\bibfnamefont {T.~W.}\ \bibnamefont
  {Kibble}},\ }\href@noop {} {\bibfield  {journal} {\bibinfo  {journal}
  {Physics Reports}\ }\textbf {\bibinfo {volume} {67}},\ \bibinfo {pages} {183}
  (\bibinfo {year} {1980})}\BibitemShut {NoStop}%
\bibitem [{\citenamefont {Zurek}(1985)}]{kzm3}%
  \BibitemOpen
  \bibfield  {author} {\bibinfo {author} {\bibfnamefont {W.~H.}\ \bibnamefont
  {Zurek}},\ }\href@noop {} {\bibfield  {journal} {\bibinfo  {journal}
  {Nature}\ }\textbf {\bibinfo {volume} {317}},\ \bibinfo {pages} {505}
  (\bibinfo {year} {1985})}\BibitemShut {NoStop}%
\bibitem [{\citenamefont {Zurek}(1993)}]{kzm4}%
  \BibitemOpen
  \bibfield  {author} {\bibinfo {author} {\bibfnamefont {W.~H.}\ \bibnamefont
  {Zurek}},\ }\href@noop {} {\bibfield  {journal} {\bibinfo  {journal} {Acta
  physica polonica. B}\ }\textbf {\bibinfo {volume} {24}},\ \bibinfo {pages}
  {1301} (\bibinfo {year} {1993})}\BibitemShut {NoStop}%
\bibitem [{\citenamefont {Zurek}(1996)}]{kzm5}%
  \BibitemOpen
  \bibfield  {author} {\bibinfo {author} {\bibfnamefont {W.~H.}\ \bibnamefont
  {Zurek}},\ }\href@noop {} {\bibfield  {journal} {\bibinfo  {journal} {Physics
  Reports}\ }\textbf {\bibinfo {volume} {276}},\ \bibinfo {pages} {177}
  (\bibinfo {year} {1996})}\BibitemShut {NoStop}%
\bibitem [{\citenamefont {Del~Campo}\ and\ \citenamefont {Zurek}(2014)}]{kzm6}%
  \BibitemOpen
  \bibfield  {author} {\bibinfo {author} {\bibfnamefont {A.}~\bibnamefont
  {Del~Campo}}\ and\ \bibinfo {author} {\bibfnamefont {W.~H.}\ \bibnamefont
  {Zurek}},\ }\href@noop {} {\bibfield  {journal} {\bibinfo  {journal}
  {International Journal of Modern Physics A}\ }\textbf {\bibinfo {volume}
  {29}},\ \bibinfo {pages} {1430018} (\bibinfo {year} {2014})}\BibitemShut
  {NoStop}%
\bibitem [{\citenamefont {Chuang}\ \emph {et~al.}(1991)\citenamefont {Chuang},
  \citenamefont {Durrer}, \citenamefont {Turok},\ and\ \citenamefont
  {Yurke}}]{issac}%
  \BibitemOpen
  \bibfield  {author} {\bibinfo {author} {\bibfnamefont {I.}~\bibnamefont
  {Chuang}}, \bibinfo {author} {\bibfnamefont {R.}~\bibnamefont {Durrer}},
  \bibinfo {author} {\bibfnamefont {N.}~\bibnamefont {Turok}}, \ and\ \bibinfo
  {author} {\bibfnamefont {B.}~\bibnamefont {Yurke}},\ }\href@noop {}
  {\bibfield  {journal} {\bibinfo  {journal} {Science}\ }\textbf {\bibinfo
  {volume} {251}},\ \bibinfo {pages} {1336} (\bibinfo {year}
  {1991})}\BibitemShut {NoStop}%
\bibitem [{\citenamefont {Lin}\ \emph {et~al.}(2014)\citenamefont {Lin},
  \citenamefont {Wang}, \citenamefont {Kamiya}, \citenamefont {Chern},
  \citenamefont {Fan}, \citenamefont {Fan}, \citenamefont {Casas},
  \citenamefont {Liu}, \citenamefont {Kiryukhin}, \citenamefont {Zurek} \emph
  {et~al.}}]{lin}%
  \BibitemOpen
  \bibfield  {author} {\bibinfo {author} {\bibfnamefont {S.-Z.}\ \bibnamefont
  {Lin}}, \bibinfo {author} {\bibfnamefont {X.}~\bibnamefont {Wang}}, \bibinfo
  {author} {\bibfnamefont {Y.}~\bibnamefont {Kamiya}}, \bibinfo {author}
  {\bibfnamefont {G.-W.}\ \bibnamefont {Chern}}, \bibinfo {author}
  {\bibfnamefont {F.}~\bibnamefont {Fan}}, \bibinfo {author} {\bibfnamefont
  {D.}~\bibnamefont {Fan}}, \bibinfo {author} {\bibfnamefont {B.}~\bibnamefont
  {Casas}}, \bibinfo {author} {\bibfnamefont {Y.}~\bibnamefont {Liu}}, \bibinfo
  {author} {\bibfnamefont {V.}~\bibnamefont {Kiryukhin}}, \bibinfo {author}
  {\bibfnamefont {W.~H.}\ \bibnamefont {Zurek}},  \emph {et~al.},\ }\href@noop
  {} {\bibfield  {journal} {\bibinfo  {journal} {Nature Physics}\ }\textbf
  {\bibinfo {volume} {10}},\ \bibinfo {pages} {970} (\bibinfo {year}
  {2014})}\BibitemShut {NoStop}%
\bibitem [{\citenamefont {Lib{\'a}l}\ \emph {et~al.}(2020)\citenamefont
  {Lib{\'a}l}, \citenamefont {Del~Campo}, \citenamefont {Nisoli}, \citenamefont
  {Reichhardt},\ and\ \citenamefont {Reichhardt}}]{spinice}%
  \BibitemOpen
  \bibfield  {author} {\bibinfo {author} {\bibfnamefont {A.}~\bibnamefont
  {Lib{\'a}l}}, \bibinfo {author} {\bibfnamefont {A.}~\bibnamefont
  {Del~Campo}}, \bibinfo {author} {\bibfnamefont {C.}~\bibnamefont {Nisoli}},
  \bibinfo {author} {\bibfnamefont {C.}~\bibnamefont {Reichhardt}}, \ and\
  \bibinfo {author} {\bibfnamefont {C.}~\bibnamefont {Reichhardt}},\
  }\href@noop {} {\bibfield  {journal} {\bibinfo  {journal} {Physical Review
  Research}\ }\textbf {\bibinfo {volume} {2}},\ \bibinfo {pages} {033433}
  (\bibinfo {year} {2020})}\BibitemShut {NoStop}%
\bibitem [{\citenamefont {Sadhasivam}\ \emph {et~al.}(2024)\citenamefont
  {Sadhasivam}, \citenamefont {Suzuki}, \citenamefont {Yan},\ and\
  \citenamefont {Sinitsyn}}]{condensed}%
  \BibitemOpen
  \bibfield  {author} {\bibinfo {author} {\bibfnamefont {V.~G.}\ \bibnamefont
  {Sadhasivam}}, \bibinfo {author} {\bibfnamefont {F.}~\bibnamefont {Suzuki}},
  \bibinfo {author} {\bibfnamefont {B.}~\bibnamefont {Yan}}, \ and\ \bibinfo
  {author} {\bibfnamefont {N.~A.}\ \bibnamefont {Sinitsyn}},\ }\href@noop {}
  {\bibfield  {journal} {\bibinfo  {journal} {Nature Communications}\ }\textbf
  {\bibinfo {volume} {15}},\ \bibinfo {pages} {10246} (\bibinfo {year}
  {2024})}\BibitemShut {NoStop}%
\bibitem [{\citenamefont {Shu}\ \emph {et~al.}(2025)\citenamefont {Shu},
  \citenamefont {Jian}, \citenamefont {Sandvik},\ and\ \citenamefont
  {Yin}}]{condensed2}%
  \BibitemOpen
  \bibfield  {author} {\bibinfo {author} {\bibfnamefont {Y.-R.}\ \bibnamefont
  {Shu}}, \bibinfo {author} {\bibfnamefont {S.-K.}\ \bibnamefont {Jian}},
  \bibinfo {author} {\bibfnamefont {A.~W.}\ \bibnamefont {Sandvik}}, \ and\
  \bibinfo {author} {\bibfnamefont {S.}~\bibnamefont {Yin}},\ }\href@noop {}
  {\bibfield  {journal} {\bibinfo  {journal} {Nature Communications}\ }\textbf
  {\bibinfo {volume} {16}},\ \bibinfo {pages} {3402} (\bibinfo {year}
  {2025})}\BibitemShut {NoStop}%
\bibitem [{\citenamefont {Schaller}\ \emph {et~al.}(2025)\citenamefont
  {Schaller}, \citenamefont {Queisser}, \citenamefont {Katoorani},
  \citenamefont {Brand}, \citenamefont {Kohlf{\"u}rst}, \citenamefont
  {Freeman}, \citenamefont {Hucht}, \citenamefont {Kratzer}, \citenamefont
  {Sothmann}, \citenamefont {Hoegen} \emph {et~al.}}]{condensed3}%
  \BibitemOpen
  \bibfield  {author} {\bibinfo {author} {\bibfnamefont {G.}~\bibnamefont
  {Schaller}}, \bibinfo {author} {\bibfnamefont {F.}~\bibnamefont {Queisser}},
  \bibinfo {author} {\bibfnamefont {S.~P.}\ \bibnamefont {Katoorani}}, \bibinfo
  {author} {\bibfnamefont {C.}~\bibnamefont {Brand}}, \bibinfo {author}
  {\bibfnamefont {C.}~\bibnamefont {Kohlf{\"u}rst}}, \bibinfo {author}
  {\bibfnamefont {M.~R.}\ \bibnamefont {Freeman}}, \bibinfo {author}
  {\bibfnamefont {A.}~\bibnamefont {Hucht}}, \bibinfo {author} {\bibfnamefont
  {P.}~\bibnamefont {Kratzer}}, \bibinfo {author} {\bibfnamefont
  {B.}~\bibnamefont {Sothmann}}, \bibinfo {author} {\bibfnamefont {M.~H.-v.}\
  \bibnamefont {Hoegen}},  \emph {et~al.},\ }\href@noop {} {\bibfield
  {journal} {\bibinfo  {journal} {Physical Review Letters}\ }\textbf {\bibinfo
  {volume} {134}},\ \bibinfo {pages} {246202} (\bibinfo {year}
  {2025})}\BibitemShut {NoStop}%
\bibitem [{\citenamefont {Balducci}\ \emph {et~al.}(2024)\citenamefont
  {Balducci}, \citenamefont {Grabarits},\ and\ \citenamefont {del
  Campo}}]{condensed4}%
  \BibitemOpen
  \bibfield  {author} {\bibinfo {author} {\bibfnamefont {F.}~\bibnamefont
  {Balducci}}, \bibinfo {author} {\bibfnamefont {A.}~\bibnamefont {Grabarits}},
  \ and\ \bibinfo {author} {\bibfnamefont {A.}~\bibnamefont {del Campo}},\
  }\href@noop {} {\bibfield  {journal} {\bibinfo  {journal} {arXiv preprint
  arXiv:2410.02520}\ } (\bibinfo {year} {2024})}\BibitemShut {NoStop}%
\bibitem [{\citenamefont {Suzuki}\ \emph {et~al.}(2025)\citenamefont {Suzuki},
  \citenamefont {Malla},\ and\ \citenamefont {Sinitsyn}}]{condensed5}%
  \BibitemOpen
  \bibfield  {author} {\bibinfo {author} {\bibfnamefont {F.}~\bibnamefont
  {Suzuki}}, \bibinfo {author} {\bibfnamefont {R.~K.}\ \bibnamefont {Malla}}, \
  and\ \bibinfo {author} {\bibfnamefont {N.~A.}\ \bibnamefont {Sinitsyn}},\
  }\href@noop {} {\bibfield  {journal} {\bibinfo  {journal} {Physical Review
  A}\ }\textbf {\bibinfo {volume} {112}},\ \bibinfo {pages} {013307} (\bibinfo
  {year} {2025})}\BibitemShut {NoStop}%
\bibitem [{\citenamefont {Du}\ \emph {et~al.}(2023)\citenamefont {Du},
  \citenamefont {Fang}, \citenamefont {Won}, \citenamefont {De}, \citenamefont
  {Huang}, \citenamefont {Xu}, \citenamefont {You}, \citenamefont
  {G{\'o}mez-Ruiz}, \citenamefont {Del~Campo},\ and\ \citenamefont
  {Cheong}}]{condensed6}%
  \BibitemOpen
  \bibfield  {author} {\bibinfo {author} {\bibfnamefont {K.}~\bibnamefont
  {Du}}, \bibinfo {author} {\bibfnamefont {X.}~\bibnamefont {Fang}}, \bibinfo
  {author} {\bibfnamefont {C.}~\bibnamefont {Won}}, \bibinfo {author}
  {\bibfnamefont {C.}~\bibnamefont {De}}, \bibinfo {author} {\bibfnamefont
  {F.-T.}\ \bibnamefont {Huang}}, \bibinfo {author} {\bibfnamefont
  {W.}~\bibnamefont {Xu}}, \bibinfo {author} {\bibfnamefont {H.}~\bibnamefont
  {You}}, \bibinfo {author} {\bibfnamefont {F.~J.}\ \bibnamefont
  {G{\'o}mez-Ruiz}}, \bibinfo {author} {\bibfnamefont {A.}~\bibnamefont
  {Del~Campo}}, \ and\ \bibinfo {author} {\bibfnamefont {S.-W.}\ \bibnamefont
  {Cheong}},\ }\href@noop {} {\bibfield  {journal} {\bibinfo  {journal} {Nature
  Physics}\ }\textbf {\bibinfo {volume} {19}},\ \bibinfo {pages} {1495}
  (\bibinfo {year} {2023})}\BibitemShut {NoStop}%
\bibitem [{\citenamefont {Damski}(2005)}]{damski2005simplest}%
  \BibitemOpen
  \bibfield  {author} {\bibinfo {author} {\bibfnamefont {B.}~\bibnamefont
  {Damski}},\ }\href@noop {} {\bibfield  {journal} {\bibinfo  {journal}
  {Physical review letters}\ }\textbf {\bibinfo {volume} {95}},\ \bibinfo
  {pages} {035701} (\bibinfo {year} {2005})}\BibitemShut {NoStop}%
\bibitem [{\citenamefont {Polkovnikov}(2005)}]{polkovnikov2005universal}%
  \BibitemOpen
  \bibfield  {author} {\bibinfo {author} {\bibfnamefont {A.}~\bibnamefont
  {Polkovnikov}},\ }\href@noop {} {\bibfield  {journal} {\bibinfo  {journal}
  {Physical Review B—Condensed Matter and Materials Physics}\ }\textbf
  {\bibinfo {volume} {72}},\ \bibinfo {pages} {161201} (\bibinfo {year}
  {2005})}\BibitemShut {NoStop}%
\bibitem [{\citenamefont {Zurek}\ \emph {et~al.}(2005)\citenamefont {Zurek},
  \citenamefont {Dorner},\ and\ \citenamefont {Zoller}}]{zurek2005dynamics}%
  \BibitemOpen
  \bibfield  {author} {\bibinfo {author} {\bibfnamefont {W.~H.}\ \bibnamefont
  {Zurek}}, \bibinfo {author} {\bibfnamefont {U.}~\bibnamefont {Dorner}}, \
  and\ \bibinfo {author} {\bibfnamefont {P.}~\bibnamefont {Zoller}},\
  }\href@noop {} {\bibfield  {journal} {\bibinfo  {journal} {Physical review
  letters}\ }\textbf {\bibinfo {volume} {95}},\ \bibinfo {pages} {105701}
  (\bibinfo {year} {2005})}\BibitemShut {NoStop}%
\bibitem [{\citenamefont {Dziarmaga}(2005)}]{dziarmaga2005dynamics}%
  \BibitemOpen
  \bibfield  {author} {\bibinfo {author} {\bibfnamefont {J.}~\bibnamefont
  {Dziarmaga}},\ }\href@noop {} {\bibfield  {journal} {\bibinfo  {journal}
  {Physical review letters}\ }\textbf {\bibinfo {volume} {95}},\ \bibinfo
  {pages} {245701} (\bibinfo {year} {2005})}\BibitemShut {NoStop}%
\bibitem [{\citenamefont {Dziarmaga}(2010)}]{dziarmaga2010dynamics}%
  \BibitemOpen
  \bibfield  {author} {\bibinfo {author} {\bibfnamefont {J.}~\bibnamefont
  {Dziarmaga}},\ }\href@noop {} {\bibfield  {journal} {\bibinfo  {journal}
  {Advances in Physics}\ }\textbf {\bibinfo {volume} {59}},\ \bibinfo {pages}
  {1063} (\bibinfo {year} {2010})}\BibitemShut {NoStop}%
\bibitem [{\citenamefont {Rams}\ \emph {et~al.}(2019)\citenamefont {Rams},
  \citenamefont {Dziarmaga},\ and\ \citenamefont {Zurek}}]{rams2019symmetry}%
  \BibitemOpen
  \bibfield  {author} {\bibinfo {author} {\bibfnamefont {M.~M.}\ \bibnamefont
  {Rams}}, \bibinfo {author} {\bibfnamefont {J.}~\bibnamefont {Dziarmaga}}, \
  and\ \bibinfo {author} {\bibfnamefont {W.~H.}\ \bibnamefont {Zurek}},\
  }\href@noop {} {\bibfield  {journal} {\bibinfo  {journal} {Physical Review
  Letters}\ }\textbf {\bibinfo {volume} {123}},\ \bibinfo {pages} {130603}
  (\bibinfo {year} {2019})}\BibitemShut {NoStop}%
\bibitem [{\citenamefont {Schmitt}\ \emph {et~al.}(2022)\citenamefont
  {Schmitt}, \citenamefont {Rams}, \citenamefont {Dziarmaga}, \citenamefont
  {Heyl},\ and\ \citenamefont {Zurek}}]{schmitt2022quantum}%
  \BibitemOpen
  \bibfield  {author} {\bibinfo {author} {\bibfnamefont {M.}~\bibnamefont
  {Schmitt}}, \bibinfo {author} {\bibfnamefont {M.~M.}\ \bibnamefont {Rams}},
  \bibinfo {author} {\bibfnamefont {J.}~\bibnamefont {Dziarmaga}}, \bibinfo
  {author} {\bibfnamefont {M.}~\bibnamefont {Heyl}}, \ and\ \bibinfo {author}
  {\bibfnamefont {W.~H.}\ \bibnamefont {Zurek}},\ }\href@noop {} {\bibfield
  {journal} {\bibinfo  {journal} {Science Advances}\ }\textbf {\bibinfo
  {volume} {8}},\ \bibinfo {pages} {eabl6850} (\bibinfo {year}
  {2022})}\BibitemShut {NoStop}%
\bibitem [{\citenamefont {Zhang}\ \emph
  {et~al.}(2025{\natexlab{a}})\citenamefont {Zhang}, \citenamefont
  {Bayocboc~Jr},\ and\ \citenamefont {Dziarmaga}}]{zhang2025kibble}%
  \BibitemOpen
  \bibfield  {author} {\bibinfo {author} {\bibfnamefont {Y.}~\bibnamefont
  {Zhang}}, \bibinfo {author} {\bibfnamefont {F.~A.}\ \bibnamefont
  {Bayocboc~Jr}}, \ and\ \bibinfo {author} {\bibfnamefont {J.}~\bibnamefont
  {Dziarmaga}},\ }\href@noop {} {\bibfield  {journal} {\bibinfo  {journal}
  {Physical Review B}\ }\textbf {\bibinfo {volume} {112}},\ \bibinfo {pages}
  {134420} (\bibinfo {year} {2025}{\natexlab{a}})}\BibitemShut {NoStop}%
\bibitem [{\citenamefont {Bayocboc~Jr}\ \emph {et~al.}(2025)\citenamefont
  {Bayocboc~Jr}, \citenamefont {Dziarmaga}, \citenamefont {Rams},\ and\
  \citenamefont {Zurek}}]{bayocboc2025persistent}%
  \BibitemOpen
  \bibfield  {author} {\bibinfo {author} {\bibfnamefont {F.~A.}\ \bibnamefont
  {Bayocboc~Jr}}, \bibinfo {author} {\bibfnamefont {J.}~\bibnamefont
  {Dziarmaga}}, \bibinfo {author} {\bibfnamefont {M.~M.}\ \bibnamefont {Rams}},
  \ and\ \bibinfo {author} {\bibfnamefont {W.~H.}\ \bibnamefont {Zurek}},\
  }\href@noop {} {\bibfield  {journal} {\bibinfo  {journal} {Physical Review
  B}\ }\textbf {\bibinfo {volume} {111}},\ \bibinfo {pages} {L100406} (\bibinfo
  {year} {2025})}\BibitemShut {NoStop}%
\bibitem [{\citenamefont {Dziarmaga}\ \emph {et~al.}(2022)\citenamefont
  {Dziarmaga}, \citenamefont {Rams},\ and\ \citenamefont
  {Zurek}}]{dziarmaga2022coherent}%
  \BibitemOpen
  \bibfield  {author} {\bibinfo {author} {\bibfnamefont {J.}~\bibnamefont
  {Dziarmaga}}, \bibinfo {author} {\bibfnamefont {M.~M.}\ \bibnamefont {Rams}},
  \ and\ \bibinfo {author} {\bibfnamefont {W.~H.}\ \bibnamefont {Zurek}},\
  }\href@noop {} {\bibfield  {journal} {\bibinfo  {journal} {Physical Review
  Letters}\ }\textbf {\bibinfo {volume} {129}},\ \bibinfo {pages} {260407}
  (\bibinfo {year} {2022})}\BibitemShut {NoStop}%
\bibitem [{\citenamefont {Gardas}\ \emph {et~al.}(2018)\citenamefont {Gardas},
  \citenamefont {Dziarmaga}, \citenamefont {Zurek},\ and\ \citenamefont
  {Zwolak}}]{gardas2018defects}%
  \BibitemOpen
  \bibfield  {author} {\bibinfo {author} {\bibfnamefont {B.}~\bibnamefont
  {Gardas}}, \bibinfo {author} {\bibfnamefont {J.}~\bibnamefont {Dziarmaga}},
  \bibinfo {author} {\bibfnamefont {W.~H.}\ \bibnamefont {Zurek}}, \ and\
  \bibinfo {author} {\bibfnamefont {M.}~\bibnamefont {Zwolak}},\ }\href@noop {}
  {\bibfield  {journal} {\bibinfo  {journal} {Scientific reports}\ }\textbf
  {\bibinfo {volume} {8}},\ \bibinfo {pages} {4539} (\bibinfo {year}
  {2018})}\BibitemShut {NoStop}%
\bibitem [{\citenamefont {Keesling}\ \emph {et~al.}(2019)\citenamefont
  {Keesling}, \citenamefont {Omran}, \citenamefont {Levine}, \citenamefont
  {Bernien}, \citenamefont {Pichler}, \citenamefont {Choi}, \citenamefont
  {Samajdar}, \citenamefont {Schwartz}, \citenamefont {Silvi}, \citenamefont
  {Sachdev} \emph {et~al.}}]{qc3}%
  \BibitemOpen
  \bibfield  {author} {\bibinfo {author} {\bibfnamefont {A.}~\bibnamefont
  {Keesling}}, \bibinfo {author} {\bibfnamefont {A.}~\bibnamefont {Omran}},
  \bibinfo {author} {\bibfnamefont {H.}~\bibnamefont {Levine}}, \bibinfo
  {author} {\bibfnamefont {H.}~\bibnamefont {Bernien}}, \bibinfo {author}
  {\bibfnamefont {H.}~\bibnamefont {Pichler}}, \bibinfo {author} {\bibfnamefont
  {S.}~\bibnamefont {Choi}}, \bibinfo {author} {\bibfnamefont {R.}~\bibnamefont
  {Samajdar}}, \bibinfo {author} {\bibfnamefont {S.}~\bibnamefont {Schwartz}},
  \bibinfo {author} {\bibfnamefont {P.}~\bibnamefont {Silvi}}, \bibinfo
  {author} {\bibfnamefont {S.}~\bibnamefont {Sachdev}},  \emph {et~al.},\
  }\href@noop {} {\bibfield  {journal} {\bibinfo  {journal} {Nature}\ }\textbf
  {\bibinfo {volume} {568}},\ \bibinfo {pages} {207} (\bibinfo {year}
  {2019})}\BibitemShut {NoStop}%
\bibitem [{\citenamefont {Zhang}\ \emph
  {et~al.}(2025{\natexlab{b}})\citenamefont {Zhang}, \citenamefont
  {Bayocboc~Jr},\ and\ \citenamefont {Dziarmaga}}]{qc}%
  \BibitemOpen
  \bibfield  {author} {\bibinfo {author} {\bibfnamefont {Y.}~\bibnamefont
  {Zhang}}, \bibinfo {author} {\bibfnamefont {F.~A.}\ \bibnamefont
  {Bayocboc~Jr}}, \ and\ \bibinfo {author} {\bibfnamefont {J.}~\bibnamefont
  {Dziarmaga}},\ }\href@noop {} {\bibfield  {journal} {\bibinfo  {journal}
  {arXiv preprint arXiv:2506.10771}\ } (\bibinfo {year}
  {2025}{\natexlab{b}})}\BibitemShut {NoStop}%
\bibitem [{\citenamefont {Visuri}\ \emph {et~al.}(2025)\citenamefont {Visuri},
  \citenamefont {Cadavid}, \citenamefont {Bhargava}, \citenamefont {Romero},
  \citenamefont {Grabarits}, \citenamefont {Chandarana}, \citenamefont
  {Solano}, \citenamefont {del Campo},\ and\ \citenamefont {Hegade}}]{qc2}%
  \BibitemOpen
  \bibfield  {author} {\bibinfo {author} {\bibfnamefont {A.-M.}\ \bibnamefont
  {Visuri}}, \bibinfo {author} {\bibfnamefont {A.~G.}\ \bibnamefont {Cadavid}},
  \bibinfo {author} {\bibfnamefont {B.~A.}\ \bibnamefont {Bhargava}}, \bibinfo
  {author} {\bibfnamefont {S.~V.}\ \bibnamefont {Romero}}, \bibinfo {author}
  {\bibfnamefont {A.}~\bibnamefont {Grabarits}}, \bibinfo {author}
  {\bibfnamefont {P.}~\bibnamefont {Chandarana}}, \bibinfo {author}
  {\bibfnamefont {E.}~\bibnamefont {Solano}}, \bibinfo {author} {\bibfnamefont
  {A.}~\bibnamefont {del Campo}}, \ and\ \bibinfo {author} {\bibfnamefont
  {N.~N.}\ \bibnamefont {Hegade}},\ }\href@noop {} {\bibfield  {journal}
  {\bibinfo  {journal} {arXiv preprint arXiv:2502.15100}\ } (\bibinfo {year}
  {2025})}\BibitemShut {NoStop}%
\bibitem [{\citenamefont {Sch{\"u}tzhold}(2025)}]{qc4}%
  \BibitemOpen
  \bibfield  {author} {\bibinfo {author} {\bibfnamefont {R.}~\bibnamefont
  {Sch{\"u}tzhold}},\ }\href@noop {} {\bibfield  {journal} {\bibinfo  {journal}
  {Progress in Particle and Nuclear Physics}\ ,\ \bibinfo {pages} {104198}}
  (\bibinfo {year} {2025})}\BibitemShut {NoStop}%
\bibitem [{\citenamefont {Xu}\ \emph {et~al.}(2018)\citenamefont {Xu},
  \citenamefont {Castelnovo}, \citenamefont {Melko}, \citenamefont {Chamon},\
  and\ \citenamefont {Sandvik}}]{xu2018dynamic}%
  \BibitemOpen
  \bibfield  {author} {\bibinfo {author} {\bibfnamefont {N.}~\bibnamefont
  {Xu}}, \bibinfo {author} {\bibfnamefont {C.}~\bibnamefont {Castelnovo}},
  \bibinfo {author} {\bibfnamefont {R.~G.}\ \bibnamefont {Melko}}, \bibinfo
  {author} {\bibfnamefont {C.}~\bibnamefont {Chamon}}, \ and\ \bibinfo {author}
  {\bibfnamefont {A.~W.}\ \bibnamefont {Sandvik}},\ }\href@noop {} {\bibfield
  {journal} {\bibinfo  {journal} {Physical Review B}\ }\textbf {\bibinfo
  {volume} {97}},\ \bibinfo {pages} {024432} (\bibinfo {year}
  {2018})}\BibitemShut {NoStop}%
\bibitem [{\citenamefont {Suzuki}\ and\ \citenamefont
  {Zurek}(2025)}]{suzuki2025deconstructing}%
  \BibitemOpen
  \bibfield  {author} {\bibinfo {author} {\bibfnamefont {F.}~\bibnamefont
  {Suzuki}}\ and\ \bibinfo {author} {\bibfnamefont {W.~H.}\ \bibnamefont
  {Zurek}},\ }\href@noop {} {\bibfield  {journal} {\bibinfo  {journal}
  {Proceedings of the National Academy of Sciences}\ }\textbf {\bibinfo
  {volume} {122}},\ \bibinfo {pages} {e2523903122} (\bibinfo {year}
  {2025})}\BibitemShut {NoStop}%
\bibitem [{\citenamefont {Carleo}\ \emph {et~al.}(2019)\citenamefont {Carleo},
  \citenamefont {Cirac}, \citenamefont {Cranmer}, \citenamefont {Daudet},
  \citenamefont {Schuld}, \citenamefont {Tishby}, \citenamefont
  {Vogt-Maranto},\ and\ \citenamefont {Zdeborov{\'a}}}]{ML}%
  \BibitemOpen
  \bibfield  {author} {\bibinfo {author} {\bibfnamefont {G.}~\bibnamefont
  {Carleo}}, \bibinfo {author} {\bibfnamefont {I.}~\bibnamefont {Cirac}},
  \bibinfo {author} {\bibfnamefont {K.}~\bibnamefont {Cranmer}}, \bibinfo
  {author} {\bibfnamefont {L.}~\bibnamefont {Daudet}}, \bibinfo {author}
  {\bibfnamefont {M.}~\bibnamefont {Schuld}}, \bibinfo {author} {\bibfnamefont
  {N.}~\bibnamefont {Tishby}}, \bibinfo {author} {\bibfnamefont
  {L.}~\bibnamefont {Vogt-Maranto}}, \ and\ \bibinfo {author} {\bibfnamefont
  {L.}~\bibnamefont {Zdeborov{\'a}}},\ }\href@noop {} {\bibfield  {journal}
  {\bibinfo  {journal} {Reviews of Modern Physics}\ }\textbf {\bibinfo {volume}
  {91}},\ \bibinfo {pages} {045002} (\bibinfo {year} {2019})}\BibitemShut
  {NoStop}%
\bibitem [{\citenamefont {Carrasquilla}\ and\ \citenamefont
  {Melko}(2017)}]{phaseML}%
  \BibitemOpen
  \bibfield  {author} {\bibinfo {author} {\bibfnamefont {J.}~\bibnamefont
  {Carrasquilla}}\ and\ \bibinfo {author} {\bibfnamefont {R.~G.}\ \bibnamefont
  {Melko}},\ }\href@noop {} {\bibfield  {journal} {\bibinfo  {journal} {Nature
  Physics}\ }\textbf {\bibinfo {volume} {13}},\ \bibinfo {pages} {431}
  (\bibinfo {year} {2017})}\BibitemShut {NoStop}%
\bibitem [{\citenamefont {Van~Nieuwenburg}\ \emph {et~al.}(2017)\citenamefont
  {Van~Nieuwenburg}, \citenamefont {Liu},\ and\ \citenamefont
  {Huber}}]{phaseML2}%
  \BibitemOpen
  \bibfield  {author} {\bibinfo {author} {\bibfnamefont {E.~P.}\ \bibnamefont
  {Van~Nieuwenburg}}, \bibinfo {author} {\bibfnamefont {Y.-H.}\ \bibnamefont
  {Liu}}, \ and\ \bibinfo {author} {\bibfnamefont {S.~D.}\ \bibnamefont
  {Huber}},\ }\href@noop {} {\bibfield  {journal} {\bibinfo  {journal} {Nature
  Physics}\ }\textbf {\bibinfo {volume} {13}},\ \bibinfo {pages} {435}
  (\bibinfo {year} {2017})}\BibitemShut {NoStop}%
\bibitem [{\citenamefont {Wang}(2016)}]{phaseML3}%
  \BibitemOpen
  \bibfield  {author} {\bibinfo {author} {\bibfnamefont {L.}~\bibnamefont
  {Wang}},\ }\href@noop {} {\bibfield  {journal} {\bibinfo  {journal} {Physical
  Review B}\ }\textbf {\bibinfo {volume} {94}},\ \bibinfo {pages} {195105}
  (\bibinfo {year} {2016})}\BibitemShut {NoStop}%
\bibitem [{\citenamefont {McDermott}\ \emph {et~al.}(2020)\citenamefont
  {McDermott}, \citenamefont {Reichhardt},\ and\ \citenamefont
  {Reichhardt}}]{phaseML4}%
  \BibitemOpen
  \bibfield  {author} {\bibinfo {author} {\bibfnamefont {D.}~\bibnamefont
  {McDermott}}, \bibinfo {author} {\bibfnamefont {C.}~\bibnamefont
  {Reichhardt}}, \ and\ \bibinfo {author} {\bibfnamefont {C.}~\bibnamefont
  {Reichhardt}},\ }\href@noop {} {\bibfield  {journal} {\bibinfo  {journal}
  {Physical Review E}\ }\textbf {\bibinfo {volume} {101}},\ \bibinfo {pages}
  {042101} (\bibinfo {year} {2020})}\BibitemShut {NoStop}%
\bibitem [{\citenamefont {Tang}\ \emph {et~al.}(2024)\citenamefont {Tang},
  \citenamefont {Liu}, \citenamefont {Zhang},\ and\ \citenamefont
  {Zhang}}]{phaseML5}%
  \BibitemOpen
  \bibfield  {author} {\bibinfo {author} {\bibfnamefont {Y.}~\bibnamefont
  {Tang}}, \bibinfo {author} {\bibfnamefont {J.}~\bibnamefont {Liu}}, \bibinfo
  {author} {\bibfnamefont {J.}~\bibnamefont {Zhang}}, \ and\ \bibinfo {author}
  {\bibfnamefont {P.}~\bibnamefont {Zhang}},\ }\href@noop {} {\bibfield
  {journal} {\bibinfo  {journal} {Nature Communications}\ }\textbf {\bibinfo
  {volume} {15}},\ \bibinfo {pages} {1117} (\bibinfo {year}
  {2024})}\BibitemShut {NoStop}%
\bibitem [{\citenamefont {Beach}\ \emph {et~al.}(2018)\citenamefont {Beach},
  \citenamefont {Golubeva},\ and\ \citenamefont {Melko}}]{phaseML6}%
  \BibitemOpen
  \bibfield  {author} {\bibinfo {author} {\bibfnamefont {M.~J.}\ \bibnamefont
  {Beach}}, \bibinfo {author} {\bibfnamefont {A.}~\bibnamefont {Golubeva}}, \
  and\ \bibinfo {author} {\bibfnamefont {R.~G.}\ \bibnamefont {Melko}},\
  }\href@noop {} {\bibfield  {journal} {\bibinfo  {journal} {Physical Review
  B}\ }\textbf {\bibinfo {volume} {97}},\ \bibinfo {pages} {045207} (\bibinfo
  {year} {2018})}\BibitemShut {NoStop}%
\bibitem [{\citenamefont {Rodriguez-Nieva}\ and\ \citenamefont
  {Scheurer}(2019)}]{phaseML7}%
  \BibitemOpen
  \bibfield  {author} {\bibinfo {author} {\bibfnamefont {J.~F.}\ \bibnamefont
  {Rodriguez-Nieva}}\ and\ \bibinfo {author} {\bibfnamefont {M.~S.}\
  \bibnamefont {Scheurer}},\ }\href@noop {} {\bibfield  {journal} {\bibinfo
  {journal} {Nature Physics}\ }\textbf {\bibinfo {volume} {15}},\ \bibinfo
  {pages} {790} (\bibinfo {year} {2019})}\BibitemShut {NoStop}%
\bibitem [{\citenamefont {Zhang}\ \emph {et~al.}(2018)\citenamefont {Zhang},
  \citenamefont {Shen},\ and\ \citenamefont {Zhai}}]{phaseML8}%
  \BibitemOpen
  \bibfield  {author} {\bibinfo {author} {\bibfnamefont {P.}~\bibnamefont
  {Zhang}}, \bibinfo {author} {\bibfnamefont {H.}~\bibnamefont {Shen}}, \ and\
  \bibinfo {author} {\bibfnamefont {H.}~\bibnamefont {Zhai}},\ }\href@noop {}
  {\bibfield  {journal} {\bibinfo  {journal} {Physical review letters}\
  }\textbf {\bibinfo {volume} {120}},\ \bibinfo {pages} {066401} (\bibinfo
  {year} {2018})}\BibitemShut {NoStop}%
\bibitem [{\citenamefont {Vargas-Hern{\'a}ndez}\ \emph
  {et~al.}(2018)\citenamefont {Vargas-Hern{\'a}ndez}, \citenamefont {Sous},
  \citenamefont {Berciu},\ and\ \citenamefont {Krems}}]{phaseML9}%
  \BibitemOpen
  \bibfield  {author} {\bibinfo {author} {\bibfnamefont {R.~A.}\ \bibnamefont
  {Vargas-Hern{\'a}ndez}}, \bibinfo {author} {\bibfnamefont {J.}~\bibnamefont
  {Sous}}, \bibinfo {author} {\bibfnamefont {M.}~\bibnamefont {Berciu}}, \ and\
  \bibinfo {author} {\bibfnamefont {R.~V.}\ \bibnamefont {Krems}},\ }\href@noop
  {} {\bibfield  {journal} {\bibinfo  {journal} {Physical review letters}\
  }\textbf {\bibinfo {volume} {121}},\ \bibinfo {pages} {255702} (\bibinfo
  {year} {2018})}\BibitemShut {NoStop}%
\bibitem [{\citenamefont {Miles}\ \emph {et~al.}(2023)\citenamefont {Miles},
  \citenamefont {Samajdar}, \citenamefont {Ebadi}, \citenamefont {Wang},
  \citenamefont {Pichler}, \citenamefont {Sachdev}, \citenamefont {Lukin},
  \citenamefont {Greiner}, \citenamefont {Weinberger},\ and\ \citenamefont
  {Kim}}]{quantumML}%
  \BibitemOpen
  \bibfield  {author} {\bibinfo {author} {\bibfnamefont {C.}~\bibnamefont
  {Miles}}, \bibinfo {author} {\bibfnamefont {R.}~\bibnamefont {Samajdar}},
  \bibinfo {author} {\bibfnamefont {S.}~\bibnamefont {Ebadi}}, \bibinfo
  {author} {\bibfnamefont {T.~T.}\ \bibnamefont {Wang}}, \bibinfo {author}
  {\bibfnamefont {H.}~\bibnamefont {Pichler}}, \bibinfo {author} {\bibfnamefont
  {S.}~\bibnamefont {Sachdev}}, \bibinfo {author} {\bibfnamefont {M.~D.}\
  \bibnamefont {Lukin}}, \bibinfo {author} {\bibfnamefont {M.}~\bibnamefont
  {Greiner}}, \bibinfo {author} {\bibfnamefont {K.~Q.}\ \bibnamefont
  {Weinberger}}, \ and\ \bibinfo {author} {\bibfnamefont {E.-A.}\ \bibnamefont
  {Kim}},\ }\href@noop {} {\bibfield  {journal} {\bibinfo  {journal} {Physical
  Review Research}\ }\textbf {\bibinfo {volume} {5}},\ \bibinfo {pages}
  {013026} (\bibinfo {year} {2023})}\BibitemShut {NoStop}%
\bibitem [{\citenamefont {Biamonte}\ \emph {et~al.}(2017)\citenamefont
  {Biamonte}, \citenamefont {Wittek}, \citenamefont {Pancotti}, \citenamefont
  {Rebentrost}, \citenamefont {Wiebe},\ and\ \citenamefont
  {Lloyd}}]{quantumML2}%
  \BibitemOpen
  \bibfield  {author} {\bibinfo {author} {\bibfnamefont {J.}~\bibnamefont
  {Biamonte}}, \bibinfo {author} {\bibfnamefont {P.}~\bibnamefont {Wittek}},
  \bibinfo {author} {\bibfnamefont {N.}~\bibnamefont {Pancotti}}, \bibinfo
  {author} {\bibfnamefont {P.}~\bibnamefont {Rebentrost}}, \bibinfo {author}
  {\bibfnamefont {N.}~\bibnamefont {Wiebe}}, \ and\ \bibinfo {author}
  {\bibfnamefont {S.}~\bibnamefont {Lloyd}},\ }\href@noop {} {\bibfield
  {journal} {\bibinfo  {journal} {Nature}\ }\textbf {\bibinfo {volume} {549}},\
  \bibinfo {pages} {195} (\bibinfo {year} {2017})}\BibitemShut {NoStop}%
\bibitem [{\citenamefont {Dong}\ \emph {et~al.}(2019)\citenamefont {Dong},
  \citenamefont {Pollmann},\ and\ \citenamefont {Zhang}}]{quantumML3}%
  \BibitemOpen
  \bibfield  {author} {\bibinfo {author} {\bibfnamefont {X.-Y.}\ \bibnamefont
  {Dong}}, \bibinfo {author} {\bibfnamefont {F.}~\bibnamefont {Pollmann}}, \
  and\ \bibinfo {author} {\bibfnamefont {X.-F.}\ \bibnamefont {Zhang}},\
  }\href@noop {} {\bibfield  {journal} {\bibinfo  {journal} {Physical Review
  B}\ }\textbf {\bibinfo {volume} {99}},\ \bibinfo {pages} {121104} (\bibinfo
  {year} {2019})}\BibitemShut {NoStop}%
\bibitem [{\citenamefont {Scheurer}\ and\ \citenamefont
  {Slager}(2020)}]{scheurer2020unsupervised}%
  \BibitemOpen
  \bibfield  {author} {\bibinfo {author} {\bibfnamefont {M.~S.}\ \bibnamefont
  {Scheurer}}\ and\ \bibinfo {author} {\bibfnamefont {R.-J.}\ \bibnamefont
  {Slager}},\ }\href@noop {} {\bibfield  {journal} {\bibinfo  {journal}
  {Physical review letters}\ }\textbf {\bibinfo {volume} {124}},\ \bibinfo
  {pages} {226401} (\bibinfo {year} {2020})}\BibitemShut {NoStop}%
\bibitem [{\citenamefont {K{\"a}ming}\ \emph {et~al.}(2021)\citenamefont
  {K{\"a}ming}, \citenamefont {Dawid}, \citenamefont {Kottmann}, \citenamefont
  {Lewenstein}, \citenamefont {Sengstock}, \citenamefont {Dauphin},\ and\
  \citenamefont {Weitenberg}}]{kaming2021unsupervised}%
  \BibitemOpen
  \bibfield  {author} {\bibinfo {author} {\bibfnamefont {N.}~\bibnamefont
  {K{\"a}ming}}, \bibinfo {author} {\bibfnamefont {A.}~\bibnamefont {Dawid}},
  \bibinfo {author} {\bibfnamefont {K.}~\bibnamefont {Kottmann}}, \bibinfo
  {author} {\bibfnamefont {M.}~\bibnamefont {Lewenstein}}, \bibinfo {author}
  {\bibfnamefont {K.}~\bibnamefont {Sengstock}}, \bibinfo {author}
  {\bibfnamefont {A.}~\bibnamefont {Dauphin}}, \ and\ \bibinfo {author}
  {\bibfnamefont {C.}~\bibnamefont {Weitenberg}},\ }\href@noop {} {\bibfield
  {journal} {\bibinfo  {journal} {Machine Learning: Science and Technology}\
  }\textbf {\bibinfo {volume} {2}},\ \bibinfo {pages} {035037} (\bibinfo {year}
  {2021})}\BibitemShut {NoStop}%
\bibitem [{\citenamefont {Che}\ \emph {et~al.}(2026)\citenamefont {Che},
  \citenamefont {Gneiting}, \citenamefont {Wang},\ and\ \citenamefont
  {Nori}}]{che2026quantum}%
  \BibitemOpen
  \bibfield  {author} {\bibinfo {author} {\bibfnamefont {Y.}~\bibnamefont
  {Che}}, \bibinfo {author} {\bibfnamefont {C.}~\bibnamefont {Gneiting}},
  \bibinfo {author} {\bibfnamefont {X.}~\bibnamefont {Wang}}, \ and\ \bibinfo
  {author} {\bibfnamefont {F.}~\bibnamefont {Nori}},\ }\href@noop {} {\bibfield
   {journal} {\bibinfo  {journal} {Nature Communications}\ }\textbf {\bibinfo
  {volume} {17}},\ \bibinfo {pages} {5179} (\bibinfo {year}
  {2026})}\BibitemShut {NoStop}%
\bibitem [{\citenamefont {Venderley}\ \emph {et~al.}(2018)\citenamefont
  {Venderley}, \citenamefont {Khemani},\ and\ \citenamefont {Kim}}]{noneq}%
  \BibitemOpen
  \bibfield  {author} {\bibinfo {author} {\bibfnamefont {J.}~\bibnamefont
  {Venderley}}, \bibinfo {author} {\bibfnamefont {V.}~\bibnamefont {Khemani}},
  \ and\ \bibinfo {author} {\bibfnamefont {E.-A.}\ \bibnamefont {Kim}},\
  }\href@noop {} {\bibfield  {journal} {\bibinfo  {journal} {Physical review
  letters}\ }\textbf {\bibinfo {volume} {120}},\ \bibinfo {pages} {257204}
  (\bibinfo {year} {2018})}\BibitemShut {NoStop}%
\bibitem [{\citenamefont {Seif}\ \emph {et~al.}(2021)\citenamefont {Seif},
  \citenamefont {Hafezi},\ and\ \citenamefont {Jarzynski}}]{noneq2}%
  \BibitemOpen
  \bibfield  {author} {\bibinfo {author} {\bibfnamefont {A.}~\bibnamefont
  {Seif}}, \bibinfo {author} {\bibfnamefont {M.}~\bibnamefont {Hafezi}}, \ and\
  \bibinfo {author} {\bibfnamefont {C.}~\bibnamefont {Jarzynski}},\ }\href@noop
  {} {\bibfield  {journal} {\bibinfo  {journal} {Nature Physics}\ }\textbf
  {\bibinfo {volume} {17}},\ \bibinfo {pages} {105} (\bibinfo {year}
  {2021})}\BibitemShut {NoStop}%
\bibitem [{\citenamefont {Boyd}\ \emph {et~al.}(2022)\citenamefont {Boyd},
  \citenamefont {Crutchfield},\ and\ \citenamefont {Gu}}]{noneq3}%
  \BibitemOpen
  \bibfield  {author} {\bibinfo {author} {\bibfnamefont {A.~B.}\ \bibnamefont
  {Boyd}}, \bibinfo {author} {\bibfnamefont {J.~P.}\ \bibnamefont
  {Crutchfield}}, \ and\ \bibinfo {author} {\bibfnamefont {M.}~\bibnamefont
  {Gu}},\ }\href@noop {} {\bibfield  {journal} {\bibinfo  {journal} {New
  Journal of Physics}\ }\textbf {\bibinfo {volume} {24}},\ \bibinfo {pages}
  {083040} (\bibinfo {year} {2022})}\BibitemShut {NoStop}%
\bibitem [{\citenamefont {Donatella}\ \emph {et~al.}(2023)\citenamefont
  {Donatella}, \citenamefont {Denis}, \citenamefont {Le~Boit{\'e}},\ and\
  \citenamefont {Ciuti}}]{donatella2023dynamics}%
  \BibitemOpen
  \bibfield  {author} {\bibinfo {author} {\bibfnamefont {K.}~\bibnamefont
  {Donatella}}, \bibinfo {author} {\bibfnamefont {Z.}~\bibnamefont {Denis}},
  \bibinfo {author} {\bibfnamefont {A.}~\bibnamefont {Le~Boit{\'e}}}, \ and\
  \bibinfo {author} {\bibfnamefont {C.}~\bibnamefont {Ciuti}},\ }\href@noop {}
  {\bibfield  {journal} {\bibinfo  {journal} {Physical Review A}\ }\textbf
  {\bibinfo {volume} {108}},\ \bibinfo {pages} {022210} (\bibinfo {year}
  {2023})}\BibitemShut {NoStop}%
\bibitem [{\citenamefont {Chowdhury}\ \emph {et~al.}(2023)\citenamefont
  {Chowdhury}, \citenamefont {Green}, \citenamefont {Park}, \citenamefont
  {Maclennan},\ and\ \citenamefont {Clark}}]{chowdhury2023topological}%
  \BibitemOpen
  \bibfield  {author} {\bibinfo {author} {\bibfnamefont {R.~A.}\ \bibnamefont
  {Chowdhury}}, \bibinfo {author} {\bibfnamefont {A.~A.}\ \bibnamefont
  {Green}}, \bibinfo {author} {\bibfnamefont {C.~S.}\ \bibnamefont {Park}},
  \bibinfo {author} {\bibfnamefont {J.~E.}\ \bibnamefont {Maclennan}}, \ and\
  \bibinfo {author} {\bibfnamefont {N.~A.}\ \bibnamefont {Clark}},\ }\href@noop
  {} {\bibfield  {journal} {\bibinfo  {journal} {Physical Review E}\ }\textbf
  {\bibinfo {volume} {107}},\ \bibinfo {pages} {044701} (\bibinfo {year}
  {2023})}\BibitemShut {NoStop}%
\bibitem [{\citenamefont {Dierking}\ \emph {et~al.}(2025)\citenamefont
  {Dierking}, \citenamefont {Moyle}, \citenamefont {Cepparulo}, \citenamefont
  {Skingle}, \citenamefont {Hern{\'a}ndez},\ and\ \citenamefont
  {Raidal}}]{dierking2025machine}%
  \BibitemOpen
  \bibfield  {author} {\bibinfo {author} {\bibfnamefont {I.}~\bibnamefont
  {Dierking}}, \bibinfo {author} {\bibfnamefont {A.}~\bibnamefont {Moyle}},
  \bibinfo {author} {\bibfnamefont {G.~M.}\ \bibnamefont {Cepparulo}}, \bibinfo
  {author} {\bibfnamefont {K.}~\bibnamefont {Skingle}}, \bibinfo {author}
  {\bibfnamefont {L.}~\bibnamefont {Hern{\'a}ndez}}, \ and\ \bibinfo {author}
  {\bibfnamefont {J.}~\bibnamefont {Raidal}},\ }\href@noop {} {\bibfield
  {journal} {\bibinfo  {journal} {Crystals}\ }\textbf {\bibinfo {volume}
  {15}},\ \bibinfo {pages} {214} (\bibinfo {year} {2025})}\BibitemShut
  {NoStop}%
\bibitem [{\citenamefont {Rumelhart}\ \emph {et~al.}(1986)\citenamefont
  {Rumelhart}, \citenamefont {Hinton},\ and\ \citenamefont {Williams}}]{RNN}%
  \BibitemOpen
  \bibfield  {author} {\bibinfo {author} {\bibfnamefont {D.~E.}\ \bibnamefont
  {Rumelhart}}, \bibinfo {author} {\bibfnamefont {G.~E.}\ \bibnamefont
  {Hinton}}, \ and\ \bibinfo {author} {\bibfnamefont {R.~J.}\ \bibnamefont
  {Williams}},\ }\href@noop {} {\bibfield  {journal} {\bibinfo  {journal}
  {nature}\ }\textbf {\bibinfo {volume} {323}},\ \bibinfo {pages} {533}
  (\bibinfo {year} {1986})}\BibitemShut {NoStop}%
\bibitem [{\citenamefont {Williams}\ and\ \citenamefont {Zipser}(1989)}]{RNN2}%
  \BibitemOpen
  \bibfield  {author} {\bibinfo {author} {\bibfnamefont {R.~J.}\ \bibnamefont
  {Williams}}\ and\ \bibinfo {author} {\bibfnamefont {D.}~\bibnamefont
  {Zipser}},\ }\href@noop {} {\bibfield  {journal} {\bibinfo  {journal} {Neural
  computation}\ }\textbf {\bibinfo {volume} {1}},\ \bibinfo {pages} {270}
  (\bibinfo {year} {1989})}\BibitemShut {NoStop}%
\bibitem [{\citenamefont {Hibat-Allah}\ \emph {et~al.}(2020)\citenamefont
  {Hibat-Allah}, \citenamefont {Ganahl}, \citenamefont {Hayward}, \citenamefont
  {Melko},\ and\ \citenamefont {Carrasquilla}}]{RNN3}%
  \BibitemOpen
  \bibfield  {author} {\bibinfo {author} {\bibfnamefont {M.}~\bibnamefont
  {Hibat-Allah}}, \bibinfo {author} {\bibfnamefont {M.}~\bibnamefont {Ganahl}},
  \bibinfo {author} {\bibfnamefont {L.~E.}\ \bibnamefont {Hayward}}, \bibinfo
  {author} {\bibfnamefont {R.~G.}\ \bibnamefont {Melko}}, \ and\ \bibinfo
  {author} {\bibfnamefont {J.}~\bibnamefont {Carrasquilla}},\ }\href@noop {}
  {\bibfield  {journal} {\bibinfo  {journal} {Physical Review Research}\
  }\textbf {\bibinfo {volume} {2}},\ \bibinfo {pages} {023358} (\bibinfo {year}
  {2020})}\BibitemShut {NoStop}%
\bibitem [{Git()}]{GitHubCode}%
  \BibitemOpen
  \href@noop {} {}\bibinfo {note} {Code available at
  \url{https://github.com/fumikasphys/topological-defect-formation/}}\BibitemShut
  {NoStop}%
\bibitem [{\citenamefont {Laguna}\ and\ \citenamefont {Zurek}(1997)}]{laguna}%
  \BibitemOpen
  \bibfield  {author} {\bibinfo {author} {\bibfnamefont {P.}~\bibnamefont
  {Laguna}}\ and\ \bibinfo {author} {\bibfnamefont {W.~H.}\ \bibnamefont
  {Zurek}},\ }\href@noop {} {\bibfield  {journal} {\bibinfo  {journal}
  {Physical Review Letters}\ }\textbf {\bibinfo {volume} {78}},\ \bibinfo
  {pages} {2519} (\bibinfo {year} {1997})}\BibitemShut {NoStop}%
\bibitem [{\citenamefont {Laguna}\ and\ \citenamefont {Zurek}(1998)}]{laguna2}%
  \BibitemOpen
  \bibfield  {author} {\bibinfo {author} {\bibfnamefont {P.}~\bibnamefont
  {Laguna}}\ and\ \bibinfo {author} {\bibfnamefont {W.~H.}\ \bibnamefont
  {Zurek}},\ }\href@noop {} {\bibfield  {journal} {\bibinfo  {journal}
  {Physical Review D}\ }\textbf {\bibinfo {volume} {58}},\ \bibinfo {pages}
  {085021} (\bibinfo {year} {1998})}\BibitemShut {NoStop}%
\bibitem [{\citenamefont {Suzuki}\ and\ \citenamefont {Zurek}(2024)}]{suzuki}%
  \BibitemOpen
  \bibfield  {author} {\bibinfo {author} {\bibfnamefont {F.}~\bibnamefont
  {Suzuki}}\ and\ \bibinfo {author} {\bibfnamefont {W.~H.}\ \bibnamefont
  {Zurek}},\ }\href@noop {} {\bibfield  {journal} {\bibinfo  {journal}
  {Physical Review Letters}\ }\textbf {\bibinfo {volume} {132}},\ \bibinfo
  {pages} {241601} (\bibinfo {year} {2024})}\BibitemShut {NoStop}%
\bibitem [{\citenamefont {Dziarmaga}\ \emph {et~al.}(1999)\citenamefont
  {Dziarmaga}, \citenamefont {Laguna},\ and\ \citenamefont {Zurek}}]{jacek}%
  \BibitemOpen
  \bibfield  {author} {\bibinfo {author} {\bibfnamefont {J.}~\bibnamefont
  {Dziarmaga}}, \bibinfo {author} {\bibfnamefont {P.}~\bibnamefont {Laguna}}, \
  and\ \bibinfo {author} {\bibfnamefont {W.~H.}\ \bibnamefont {Zurek}},\
  }\href@noop {} {\bibfield  {journal} {\bibinfo  {journal} {Physical review
  letters}\ }\textbf {\bibinfo {volume} {82}},\ \bibinfo {pages} {4749}
  (\bibinfo {year} {1999})}\BibitemShut {NoStop}%
\end{thebibliography}%
%% if required, the c\end{document}
%
% ****** End of file apssamp.tex ******

\section*{Supplemental material}

\section*{Preparation of training data}
As briefly discussed in the main text, the training data was prepared by numerically solving the Langevin equation in (1+1) dimensions for a real scalar field $\phi$, representing the order parameter evolving under the Landau-Ginzburg potential (Eq. (1)). The time evolution is initiated with $\epsilon = -1.5$ and concluded when $\epsilon$ reaches $1$. The equation is numerically solved using the fourth-order Runge-Kutta method with periodic boundary conditions on a spatial grid of $1024$ points over a length $L = 512$, corresponding to $dx = 0.5$. By repeating this process, we prepared 3000  samples that show the time evolution of the order parameter during a second-order phase transition, which are then split into a training dataset (80\%) and a validation dataset (20\%).

\section*{RNN model for predicting topological defects configuration}

RNN consisting of 256 units and a softsign activation function was used. The network was implemented using a RNN layer followed by a fully connected (dense) output layer to predict the final configuration of topological defects. The model was trained using the Adam optimizer to minimize the mean squared error (MSE) between the predicted and true $\phi$, using the short-time evolution of $\phi$ during the second-order phase transition (for $-1.5<\epsilon < 1$) as input and the corresponding final $\phi$ at $\epsilon = 1$ as the target. Training was performed for 60 epochs with a batch size of 10. The results shown in Figs. 4 (lower panel) and 5 represent averages computed over five independent runs of ML model.

Note that  the training loss measures the fit of the model to the training data during the learning process, while the validation loss  is the measure of the performance of a trained model on unknown data from the validation set. In the paper, the loss function is MSE between the predicted and true $\phi$ at the time when $t=\tau_Q$ with $\epsilon = 1$. The training and validation losses correspond to this quantity evaluated on the training and validation datasets, respectively.  An epoch (i.e., a training cycle) refers to a complete pass through the entire training dataset, during which the model is exposed to every sample and its parameters are updated based on the computed error. The dataset is partitioned into smaller subsets, known as batches, enabling the model to process portions of the training data effectively. 

\section*{Results for different values of $\theta$}

Here we show the results for values of $\theta$ different from those used in the main text. It can be seen that, although larger noise can affect the predictability of the final topological defect configuration, the main result remains unchanged: the final topological defect configuration can be predicted from inputs consisting of short intervals of the time evolution of the order parameter taken in the impulse regime, well before the freeze-out time.

\begin{figure}[h]
\centering
{%
\includegraphics[clip,width=\columnwidth]{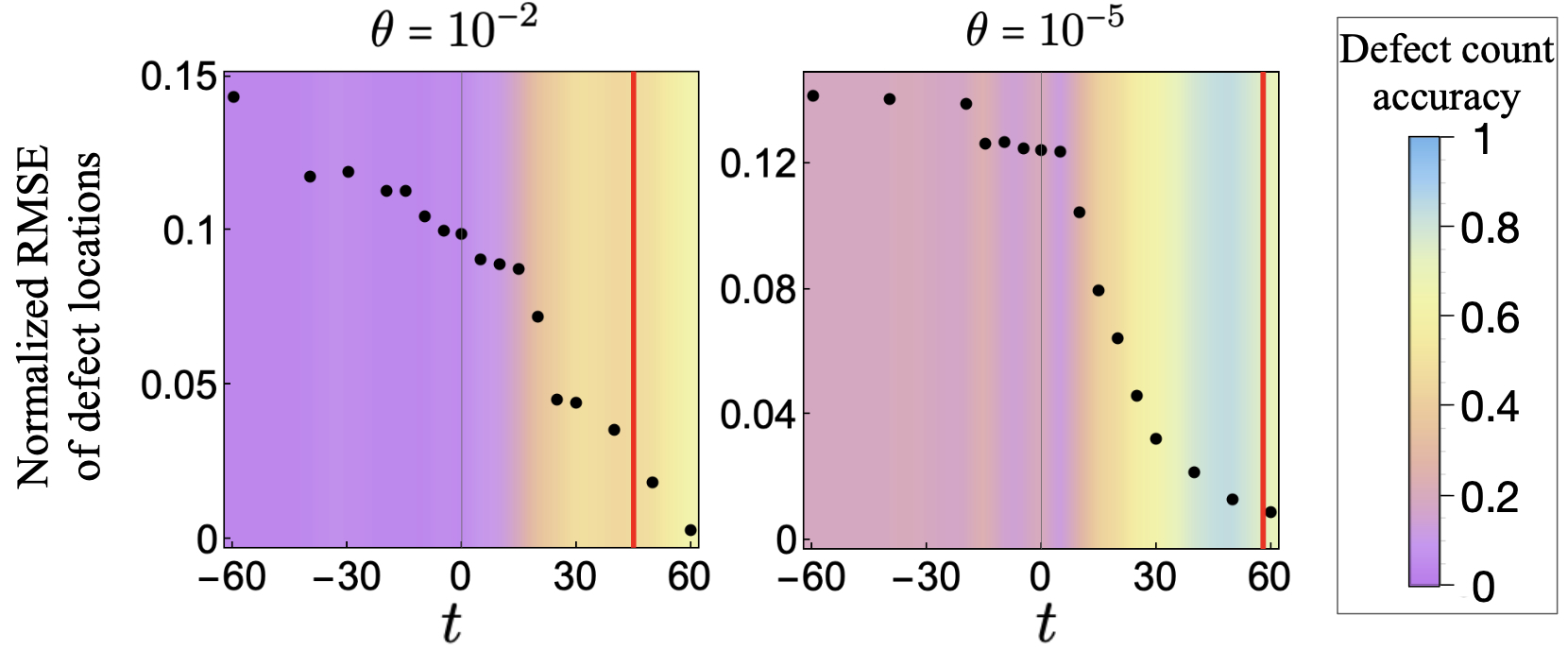} 
}
\caption{Normalized RMSE between the predicted and true defect locations for $\theta=10^{-2}$ (left) and $\theta=10^{-5}$ (right), evaluated only when the model correctly predicted the number of defects. The defect count accuracy (i.e., the fraction of cases where the number of defects is correctly predicted) is represented by the color plot.  Those are evaluated using 100 testing samples after training the ML model for 60 epochs on 3000 samples. The red vertical lines indicate the time at which $\langle |\frac{d}{dt }\phi (x,t)|\rangle$ reaches its maximum after the phase transition at $t = 0$.}
\label{figsupp}
\end{figure}

\begin{figure}
\centering
{%
\includegraphics[clip,width=\columnwidth]{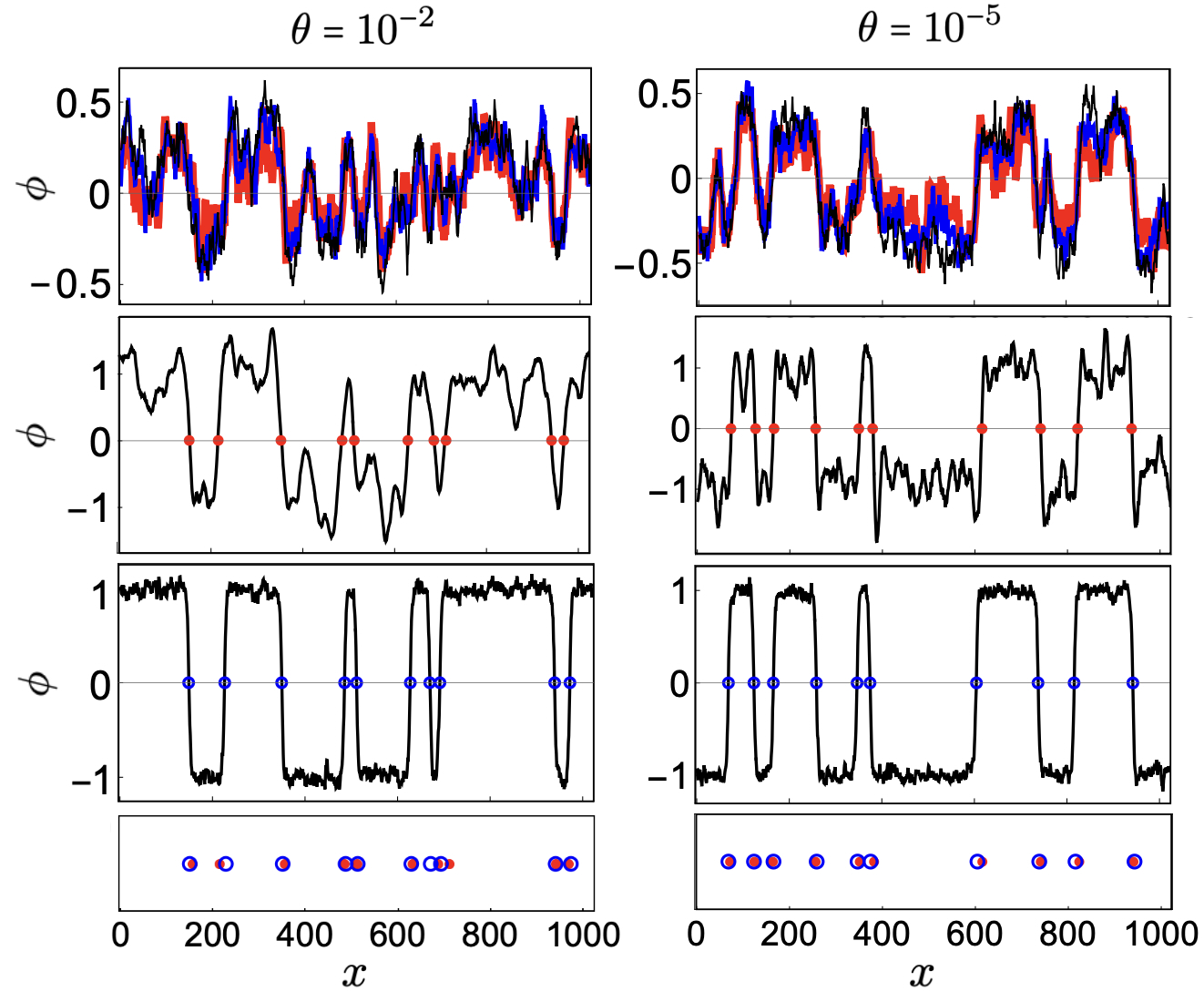} 
}
\caption{The top panels show the input data. The upper-middle panels show the predicted final topological defect configurations, while the lower-middle panels show the true final topological defect configurations. The bottom panels compare the predicted (red filled circles) and true (blue open circles) locations of the topological defects. $\theta=10^{-2}$ (left) and $\theta=10^{-5}$ (right). The input data consist of time series of $\phi$, where each series spans from $t-5$ to $t+5$ with a time step $\Delta t=1$. $t = 20$ $(\epsilon=0.16)$ for the left panel, and $t = 30$ $(\epsilon =0.23)$ for the right panel. The top panels show $\phi$ at $t-5$ ($\epsilon -0.039$; thick red line), $t$ ($\epsilon$; blue line), and $t+5$  ($\epsilon +0.039$; thin black line) as representative examples. $\tau_Q=128$.} 
\label{figsupp2}
\end{figure}

\section*{Comparison with Alternative Neural Network Architectures}

\begin{figure}
\centering
{%
\includegraphics[clip,width=0.75\columnwidth]{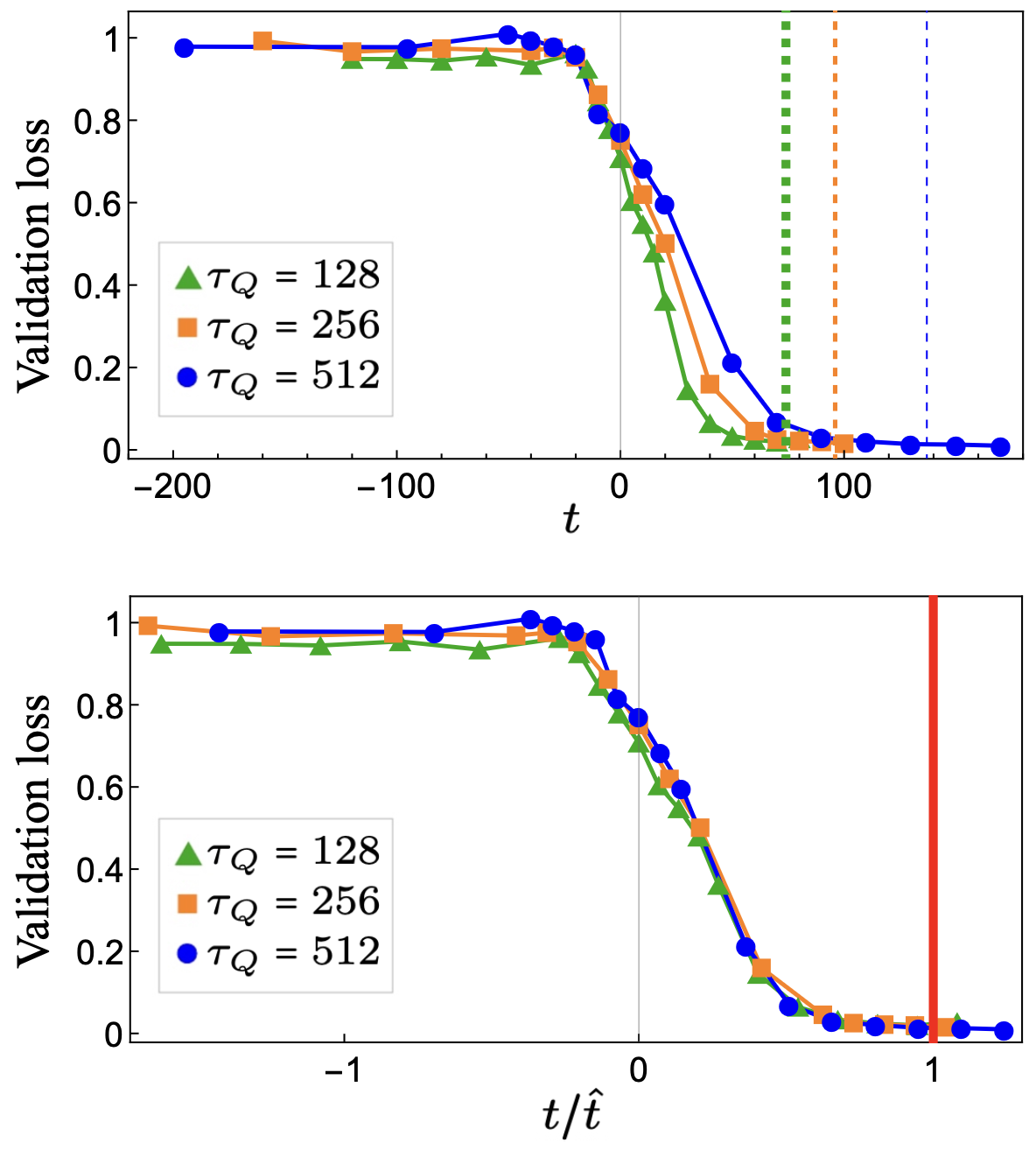} 
}
\caption{(LSTM). Validation loss for different quench timescale $\tau_Q$ achieved after 60 training epochs.  The upper panel is plotted as a function of time \( t \), while the lower panel is plotted as a function of the rescaled time $t/\hat{t}$, the KZM freeze-out time: Validation loss curves coincide after the rescaling. Vertical lines in both panels represent the corresponding freeze-out time for each \( \tau_Q \).
}
\label{LSTM}
\end{figure}

\begin{figure}
\centering
{%
\includegraphics[clip,width=0.75\columnwidth]{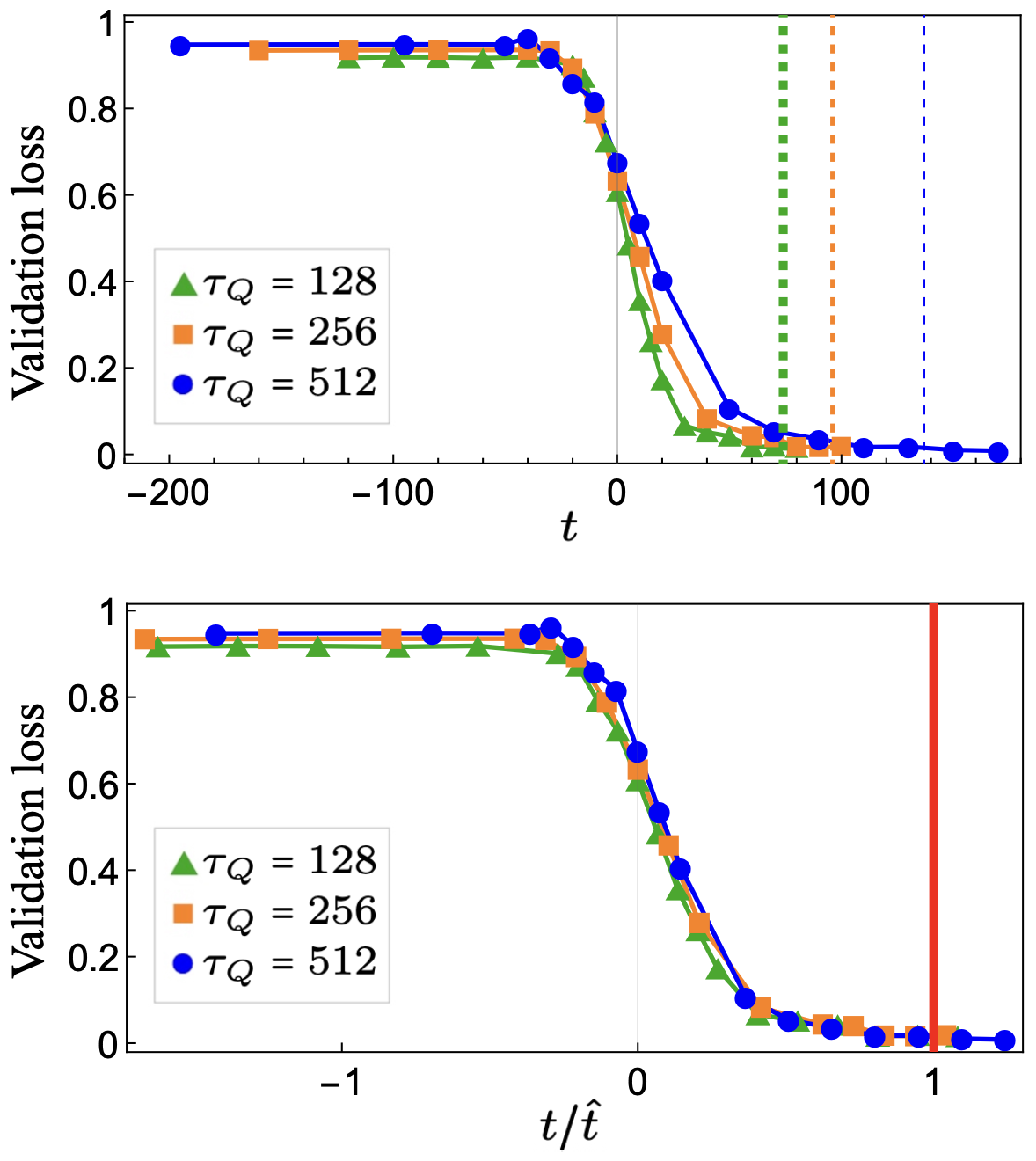} 
}
\caption{(Transformer). Validation loss for different quench timescale $\tau_Q$ achieved after 60 training epochs.  The upper panel is plotted as a function of time \( t \), while the lower panel is plotted as a function of the rescaled time $t/\hat{t}$, the KZM freeze-out time: Validation loss curves coincide after the rescaling. Vertical lines in both panels represent the corresponding freeze-out time for each \( \tau_Q \).
}
\label{transformer}
\end{figure}

To examine whether the observed prediction behavior depends on the choice of neural-network architecture, we also tested the LSTM and Transformer architectures. As shown in Figs.~S3 (obtained using an LSTM) and S4 (obtained using a Transformer), both architectures exhibit the same characteristic time dependence of the validation loss as that obtained with the RNN in Fig.~5 of the main text, including the early-time plateau and the universal collapse upon rescaling time by the KZM freeze-out time $\hat{t}$.

The agreement among these three architectures indicates that the prediction plateau and the subsequent onset of predictability are not specific to the trainability  limitations of the RNN. Although LSTMs and Transformers capture temporal correlations through mechanisms  different from those of a conventional RNN, they reproduce the same prediction behavior. This architecture independence supports the interpretation that the observed onset of predictability reflects the underlying physics of topological-defect formation governed by symmetry-breaking dynamics, rather than being an artifact of a particular neural-network architecture.

We used the same training conditions as for the RNN in the main text, except for the network architecture. The LSTM model consists of 256 units with a softsign activation function, followed by a fully connected (dense) output layer that predicts the final order-parameter configuration. The model was trained using the Adam optimizer to minimize the mean squared error (MSE) between the predicted and true $\phi$. The same dataset and training
procedure as for the RNN were used: the data were divided into training (80\%) and validation (20\%) sets, and training was performed for 60 epochs with a batch size of 10. 

For the Transformer, the recurrent layer was replaced by a self-attention architecture while the training and testing procedures were otherwise kept unchanged. The input at each time step was first projected onto a 256-dimensional representation using a fully connected layer with a softsign activation function, and a learned positional embedding was added to encode the temporal ordering of the input sequence.

The Transformer uses four-head self-attention, with a key dimension of 64 for each attention head. The attention layer is followed by a
residual connection and layer normalization. A fully connected feed-forward layer with 256 units and a softsign activation function is then applied, followed by a second residual connection and layer normalization. Finally, the Transformer output is passed through a fully connected layer to predict the final topological defect configuration.

As for the RNN and LSTM, the Transformer was trained using the Adam optimizer with the MSE loss for 60 epochs and a batch size of 10, using the same 80\%/20\% training--validation partition.

\end{document}